\documentstyle{mn}
\input{epsf.sty}
\newcommand{\figpath}{.}
\def\plottwo#1#2{\centering \leavevmode
\epsfxsize=.95\columnwidth \epsfbox{#1} \hfil
\epsfxsize=.95\columnwidth \epsfbox{#2}}
\def\plotsmall#1{\centering \leavevmode \epsfxsize=0.95\columnwidth \epsfbox{#1}}
\def\plotone#1{\centering \leavevmode \epsfxsize=1.95\columnwidth \epsfbox{#1}}
\begin{document}

\title[Star-disc encounters]{Numerical simulations of protostellar encounters \\ I. Star-disc encounters}

\author[H.M.J.~Boffin et al.]
  {H.M.J.~Boffin,
   S.J.~Watkins,
   A.S.~Bhattal,
   N.~Francis
   and A.P.~Whitworth\\
  Department of Physics and Astronomy, University of Wales, Cardiff CF2 3YB, Wales, UK.
  }

\maketitle
\begin{abstract}
It appears that most stars are born in clusters, and that at birth most stars have circumstellar discs which are comparable in size to the separations between the stars. Interactions between neighbouring stars and discs are therefore likely to play a key r\^{o}le in determining disc lifetimes, stellar masses, 
and the separations and eccentricities of binary orbits. Such interactions may also cause fragmentation of the discs, thereby triggering the formation of additional stars.

We have carried out a series of simulations of disc-star interactions using an SPH code which treats self-gravity, hydrodynamic and viscous forces. We find that interactions between discs and stars provide a mechanism for removing energy from, or adding energy to, the orbits of the stars, and for truncating the discs. However, capture during such encounters is unlikely to be an important binary formation mechanism.

A more significant consequence of such encounters is that they can trigger fragmentation of the disc, via tidally and compressionally induced gravitational instabilities, leading to the formation of additional stars. When the disc-spins and stellar orbits are randomly oriented, encounters lead to the formation of new companions to the original star in 20\% of encounters. If most encounters are prograde and coplanar, as suggested by simulations of dynamically-triggered star formation, then new companions are formed in approximately 50\% of encounters.

\end{abstract}

\section{Introduction}

In relatively quiescent molecular clouds, the star formation process may be initiated quasistatically, by ambipolar diffusion; individual protostars then  condense out indepedently of one another, via inside-out collapse. Star formation may also be initiated dynamically, and in this case multiple protostars are formed; here we have in mind impulsive triggers such as collisions between clumps in molecular clouds \cite{pringle89} or the compression of material by expanding nebula, i.e. HII regions, stellar-wind bubbles, and supernova remnants (e.g. Elmegreen \& Lada \shortcite{elmegreen:lada}). Whitworth et al. \shortcite{antisis} have argued that dynamically triggered star formation is significantly more prolific than quiescent star formation, and that it is much more likely to produce star clusters and binary systems.

In quiescent star formation, a core within a molecular cloud collapses to a protostellar disc. Viscous forces within the disc then remove angular momentum from the disc material, allowing it to accrete onto a central protostar. This accretion process continues until, by the time the protostar has reached the T-Tauri stage, most of the mass of the system is in the central star, with only a small fraction remaining in a vestigial disc. Discs are probably present around protostars for a large part of their pre-main-sequence lifetimes. 

In contrast, dynamically-triggered star formation is more chaotic. The simulations of dynamically triggered star formation reported by Turner et al. \shortcite{turner} and Whitworth et al. \shortcite{whitworth} suggest that small-$N$ clusters of protostellar discs are formed. Impulsive interactions between these discs then occur on a timescale comparable with -- or possibly even shorter than -- the timescale for the disc to accrete onto the central star. The main purpose of this paper (Paper I) and its two companion papers 
is to explore the consequences of these impulsive interactions.

Observations indicate that most stars are born in clusters (e.g. Lada et al. \shortcite{lada91}). Observations also suggest that most stars condense out of protostellar discs having radii in the range 100-1000AU, as compared to separations between stars in clusters of approximately 5000-10000AU (Strom et al.\ 1989b\nocite{strom89b}, Lada et al.\ 1991\nocite{lada91}). Therefore it seems unavoidable that most protostellar discs will, during their lifetime,  undergo at least one encounter with a neighbouring protostellar disc or star.

It has been proposed \cite{larson90} that in dense stellar clusters, capture due to gravitational drag caused by star-disc interactions could be an important binary formation mechanism. However, this mechanism cannot explain the formation of binaries with periastra ${\stackrel{<}{_{\scriptstyle \sim}}}$ 1 AU or ${\stackrel{>}{_{\scriptstyle \sim}}}$ 1000 AU; and Clarke \& Pringle \shortcite{clarke:pringle91a} have shown that, except in the densest stellar environments, capture cannot lead to a significant binary fraction.

Clarke \& Pringle \shortcite{clarke:pringle91b} and McDonald \& Clarke \shortcite{mcdonald:clarke} have modelled the evolution of an initially-bound small-$N$ cluster of stars, treating the dissipative effects of circumstellar discs with a simple prescription for the energy loss during an encounter. They find that the discs have a significant influence on the evolution of the system, leading to a considerable number of binary and multiple systems. However, their  results are sensitive to the prescription for interactions used.

For weak encounters, in which the periastron distance is much greater than the disc radius, it is possible to use linear perturbation analysis to calculate the energy and angular momentum transfer between the disc and the orbit of the two stars \cite{ostriker}. At these large separations energy and angular momentum are in general transferred from the disc to the stars. Thus long-range encounters do not provide a mechanism for capture, but can give an enhanced accretion rate within the disc.

Encounters in which the perturber passes close to, or penetrates, the disc are highly non-linear and cannot be followed analytically. It is therefore necessary to simulate such encounters numerically. Below we review the results of such simulations.

Clarke \& Pringle \shortcite{clarke:pringle93} have used a reduced three-body method to investigate the response of an accretion disc to a perturbing star. Their discs were non-self-gravitating, pressure forces were ignored, and viscous forces were treated using a pseudo-viscosity chosen to prevent fluid elements from passing through one another. They investigated three different cases. They found that for prograde, co-planar encounters, the disc is tidally stripped to about half of the periastron distance, and the perturbing star captures a significant amount of the disrupted material. In a retrograde, co-planar encounter, the disc is disrupted to about the periastron radius, and very little material is captured by the perturbing star. For orthogonal encounters, the disc is disrupted outside of periastron, and the remaining disc is twisted. The perturbing star captures almost no material.

A more detailed series of simulations has been carried out by Heller \shortcite{heller}, using smoothed particle hydrodynamics (SPH). The primary aim of this work was to calculate the tilting of the disc due to the passage of stars on non-coplanar orbits. He included pressure forces, using an adiabatic equation of state for the disc. He investigated a larger parameter space than Clarke \& Pringle, with periastron distances from 0.5 to 2 disc radii, and a range of orbital inclinations, and allowed for the effect of the disc upon the orbit of the stars.  He confined his simulations to a $1 \mbox{M}_{\sun}$ star, surrounded by a $0.1 \mbox{M}_{\sun}$ disc of radius 100 AU. This corresponds to an evolved system, rather than the massive, extended discs suggested by Lin \& Pringle \shortcite{lin:pringle90}.

Heller \shortcite{heller95} has used the same method, this time with a locally isothermal equation of state, to investigate the effect of encounters upon the disc material, and the possibility of capture as a binary formation mechanism. He finds that encounters within 2 disc radii can lead to the stripping of approximately half of the disc mass, but that capture at this epoch cannot be a dominant binary formation mechanism.

Simulations of the effect of a perturber on a disc have also been carried out using finite difference methods \cite{korycansky:papaloizou}. These simulations only treat the case of long-range, co-planar prograde encounters. They find that the encounter removes angular momentum from the disc, leading to an enhanced accretion rate, and that energy is transferred from the disc to the stars -- in agreement with the results of Ostriker \shortcite{ostriker}.

In order to discover which parts of the disc are most responsible for the transfer of energy and angular momentum, Hall, Clarke \& Pringle \shortcite{hall} have carried out a series of reduced 3-body simulations, in which hydrodynamic forces are ignored. They find that the transfers of both energy and angular momentum are usually dominated by the material that becomes unbound during the encounter. They also find that for prograde encounters, a corotation resonance causes energy and angular momentum to be transferred to the binary orbit, so that for encounters with the periastron radius greater than about twice the disc radius, the binary may become less strongly bound, or even unbound.

This paper is the first of three which describe simulations of encounters involving massive, extended protostellar discs. In this paper we describe our numerical method, discuss its limitations, and present simulations of interactions between a  discless star and a star possessing a massive, extended disc. The second and third papers (Watkins et al.\ 1997b,c) investigate interactions between two protostellar discs.

\section{Numerical method}
\label{sect:num}

The method we use in our simulations is smoothed particle hydrodynamics (SPH) combined with treecode gravity \cite{hernquist:katz}. The discs in the simulations are self-gravitating,  and experience pressure and viscous forces, as well as the gravitational attraction of the stars. The stars are represented by particles that experience only gravitational forces, and which move in the combined potential of each other and the discs. Our code uses an adaptive smoothing length $h$, so that each particle has an approximately constant number ($\sim 50$) of neighbours. For instance, the particles in the initial disc have $\bar{h} \sim \,$80AU. Gravity is kernel-softened, so $h$ is also the gravity-softening length for the particles. The stars have a constant gravity softening length of 50AU.

SPH is normally used with an artificial viscosity \cite{monaghan89}, in which the bulk viscosity and shear viscosity are coupled together.
A high bulk viscosity is necessary in order to prevent unphysical penetration of colliding particle streams, and hence to capture shocks, but this means that the resulting shear viscosity is too strong.

Here we use a prescription for the viscosity due to Watkins et al.\ \shortcite{watkins}.
This prescription allows the bulk and shear viscosities to be regulated independently, so that it is possible to have a bulk viscosity high enough to model the shocks in disc-star interactions, without having an unphysically high shear viscosity. We use a bulk viscosity $\mu_{b}=c_{s} h \rho/8$, where $c_{s}$ is the sound speed, $h$ is the SPH smoothing length and $\rho$ is the density.
This viscosity is equivalent to the bulk component of the $\alpha$-term of the SPH artificial viscosity with $\alpha \simeq 1$ \cite{murray}. For the shear viscosity we use $\mu_{s} =c_{s} h \rho/800$, which is equivalent to the $\alpha$-viscosity of Shakura \& Sunyaev \shortcite{shakura:sunyaev} with $\alpha_{\mbox{\tiny SS}} \sim 10^{-2}$. We also include the $\beta$-term from the artificial viscosity with $\beta=1$ (seeMonaghan \shortcite{monaghan89}). We have performed a variety of tests on this viscosity prescription, and they are reported in Watkins et al. \shortcite{watkins}.

Initially the numerical Reynolds number in the discs is $\sim 3000 (r/\mbox{1000AU})^{-1/2}$. There is no indication of turbulence developing, but this may simply be a consequence of the low resolution. Isolated discs evolve secularly on a viscous timescale, and in the sense predicted by Lynden-Bell \& Pringle \shortcite{lbell:pring}.

\begin{table}
\begin{tabular}{l|l}
property & value   \\ \hline
disc radius & 1000AU \\
primary disc mass & 0.5$\mbox{M}_{\sun}$ \\
primary star mass & 0.5$\mbox{M}_{\sun}$ \\
secondary star mass & 1.0$\mbox{M}_{\sun}$ \\
disc density profile & $r^{-3/2}$ \\
eccentricity of orbit & 1.0 \\
initial separation & 5000AU \\
shear viscosity & $\nu_{s} = c_{s} h /800$ \\
\end{tabular}
\caption[Physical parameters used in simulations]
{\label{table:phys}Physical parameters used in simulations}
\end{table}

\subsection{Limitations of the method}

The disc is modelled using a single layer of 2000 particles, and so it is not resolved vertically. This means that for non-coplanar encounters, the propagation of sound waves through the disc may be inhibited. However, the interactions we simulate are dominated by gravitational torques and shock compression, so sound waves have insufficient time to have a significant effect on the outcome. We infer that the thinness of the discs does not corrupt the results obtained, but this can only be confirmed by carrying out simulations  using a much larger number of particles, so as to resolve the discs vertically.

We were obliged to use a relatively small number of particles for three reasons. First, we needed to explore a rather large parameter space, and so the individual simulations had to be relatively short. Second, we needed to follow the condensation of individual protostellar discs, and this incurred a rather short time-step particularly in the later stages of the simulations. Third, we have been unable to marry our new viscosity with our multiple time-step scheme, and so the simulations become particularly compute-intensive as the new protostars condense out in the later stages.

In the central regions of the disc, where the particles are orbiting very fast, particles may find neighbours with supersonic relative velocities. Indeed, they may even find neighbours on the opposite side of the disc. This leads to the particles in the central region experiencing an unphysical shear that causes them to decelerate and spiral in towards the central star much faster than they should do. This has the effect of altering the initial surface-density profile in the immediate vicinity of the star. In combination with gravity softening, the high shear also causes the central region to form a pseudo-lattice in solid-body rotation, as can be seen in Figure 3. This problem remains isolated to within $\sim 2h$ of the centre of the disc, and does not propagate outwards. As a consequence, the inner 10\% of the disc is not correctly modelled, and any fragmentation that might occur in this region is artificially suppressed. In order to alleviate this problem, it would be necessary to use a larger number of particles, so that $h$ could be smaller and hence fewer particles would have supersonic neighbours.  The problem could also be overcome by the use of sink particles in the centre of the disc (Bate, Bonnell \& Price 1995\nocite{bate}).
\label{disc:stab}

\section{Physical and computational model}
\label{sect:model}

During the Class 0 and Class I stages, young protostars are heavily obscured by their surrounding envelopes and are thus extremely difficult to detect. The vast majority of observations of pre-main sequence stars are therefore of the later Class II and III stages, i.e. T-Tauri stars. At this stage of the evolution, the central star has accreted most of its surrounding disc, so that the discs are typically of radius 100AU and mass 0.001 to 0.01$\mbox{M}_{\sun}$ (e.g.\ Beckwith \& Sargent 1993\nocite{beckwith:sargent}). However, observations indicate that by the time stars reach the T-Tauri stage, they already have a binarity at least equal to that of main sequence stars \cite{mathieu94}. Whatever determines the binary statistics of protostars must therefore occur at an earlier stage than the T-Tauri phase. For this reason, we have chosen to model the case in which the star and disc masses are equal, and the disc is extended (1000AU). This corresponds to a Class 0 or young Class I object. The primary star and its disc are both of mass 0.5$\mbox{M}_{\sun}$, whilst for the star-disc encounters, the perturbing star has mass 1.0$\mbox{M}_{\sun}$, equal to the total mass of the system it is encountering. The physical parameters used in this paper are summarised in Table \ref{table:phys}. 

Lin \& Pringle \shortcite{lin:pringle90} found in their collapse calculations that when a disc formed, its surface density profile began as $\Sigma \propto r^{-1}$, and evolved to $\Sigma \propto r^{-3/2}$. This evolution was driven by transport of angular momentum due to gravitational instabilities in the disc. The simulations presented here therefore use an initial surface density profile $\Sigma \propto (r+r_{0})^{-3/2}$, where $r_{0}=0.02 r_{disc}$ is a small term to smooth out the central singularity. In the standard simulations, the disc is evolved in isolation before the start of the encounter, in order to remove any initial transients, and to allow the disc to reach a state in which it evolves on a secular, rather than dynamic, timescale.

The discs are taken to have an isothermal equation of state at 10K. As the discs are optically thin to their mm cooling radiation, this should be a good approximation except in the very central regions, where radiation released from Kelvin-Helmholtz contraction of the central protostar will heat the disc. At densities of greater than $10^{-13} \mbox{ g cm}^{-3}$, condensation is slowed down by making the equation of state polytropic with exponent 1.7. The results of the simulations can be rescaled to different physical units. For example, they are also applicable to 300AU radius discs which have a temperature of 33K.

The orbit of the perturbing star is expected to be mildly hyperbolic (Larson 1990). There is very little difference energetically between mildly hyperbolic  encounters and parabolic ones, so for simplicity all of the encounters modelled here are parabolic. We have carried out a suite of standard simulations varying the periastron distance, $r_{peri}$, and the orbital inclination, $\phi$. The simulations carried out are listed in Table \ref{runs}. We begin the simulations with the perturbing star at five times the disc radius.

\begin{figure*}
\plottwo{\figpath/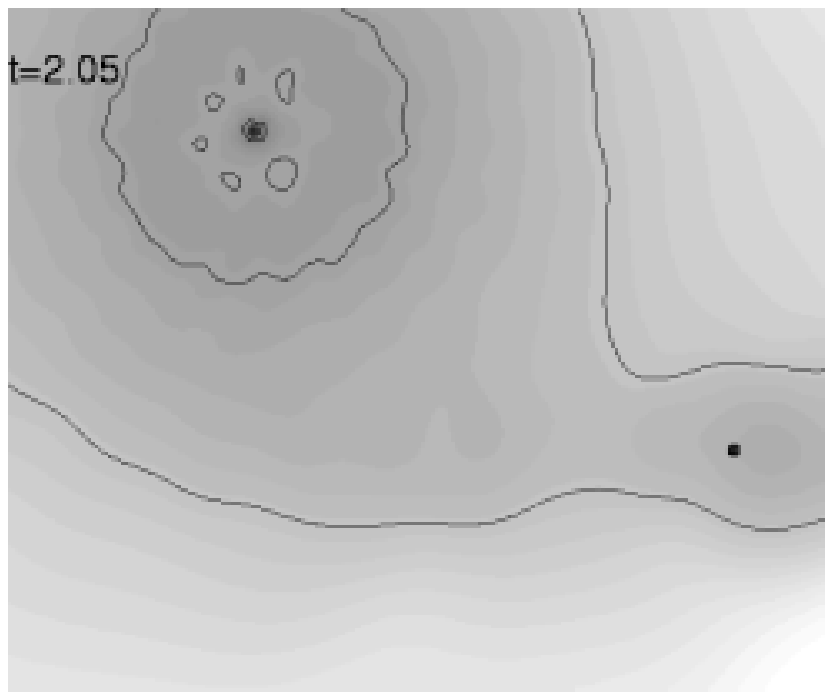}{\figpath/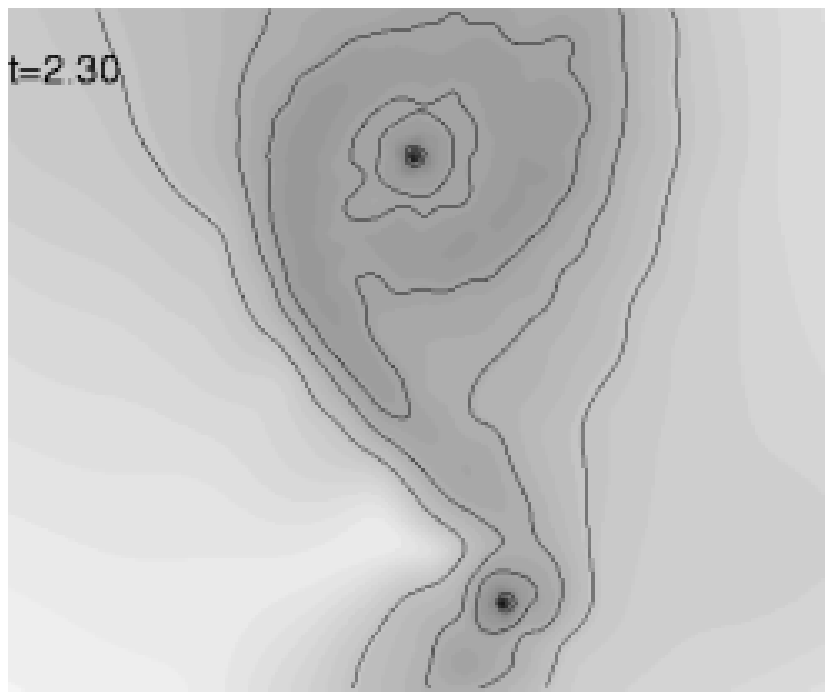}\\
\vspace{6pt}
\plottwo{\figpath/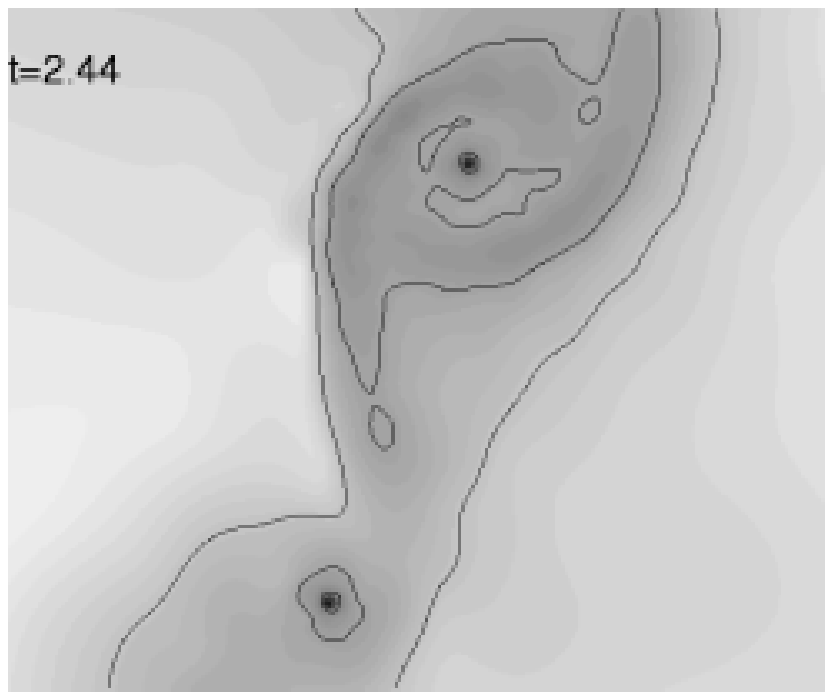}{\figpath/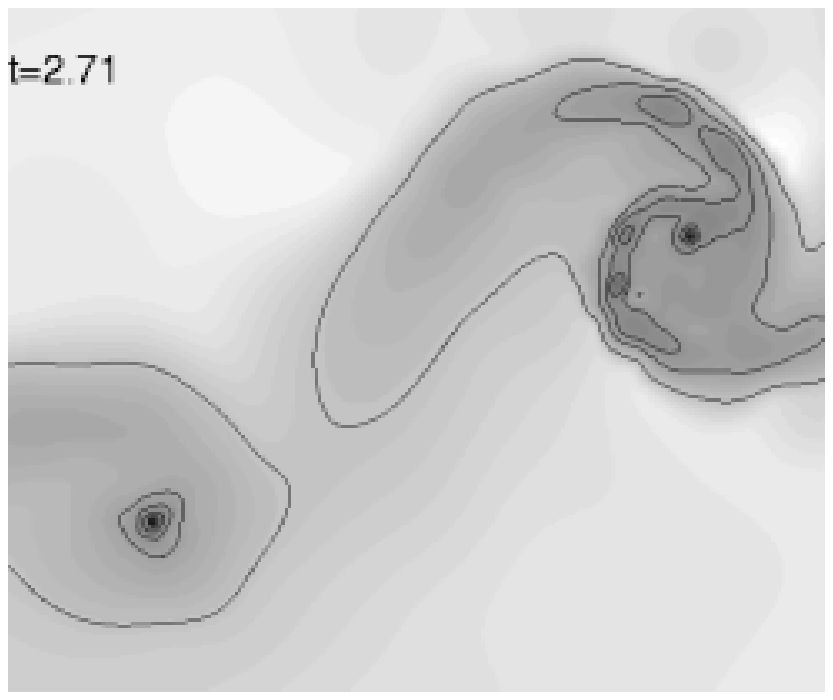}\\
\vspace{6pt}
\plottwo{\figpath/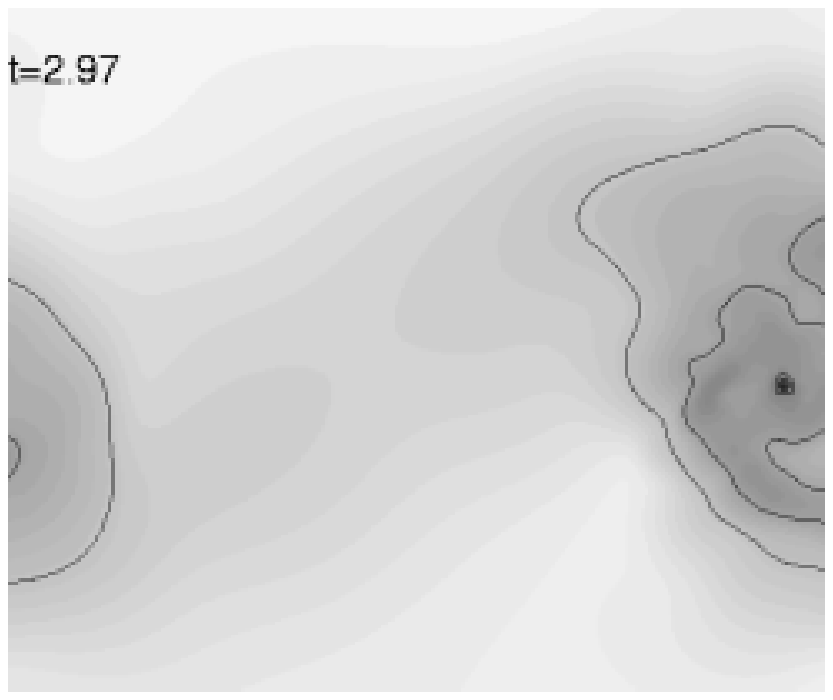}{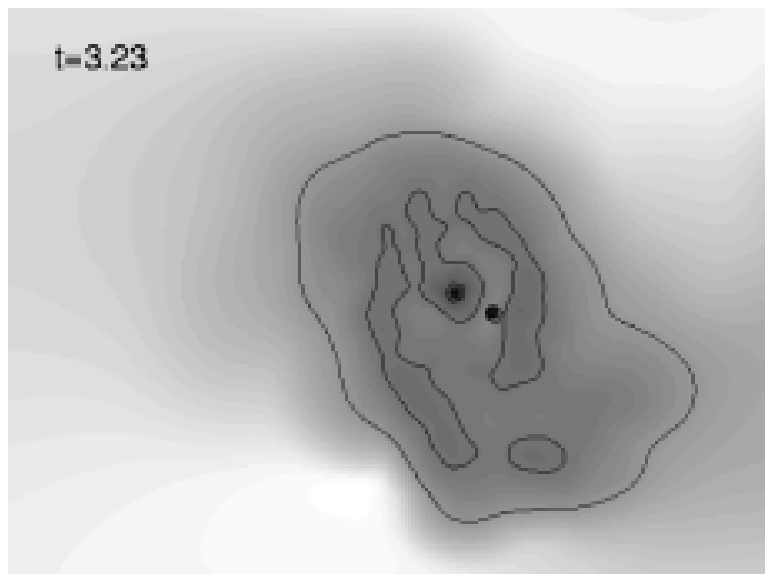}\\
\caption[Simulation ds02 : $\phi=0$, $r_{peri}=1000 \mbox{AU}$. 2000 $\times$ 1500 AU region.]{\label{fig:ds02}Simulation ds02 : $\phi=0$, $r_{peri}=1000 \mbox{AU}$. 2000 $\times$ 1500 AU region. Contour levels at [0.3, 1 (frame 2), 3, 9 (frame 4), 30, 300] $\mbox{g cm}^{-2}$. The perturber approaches from the right-hand side. It accretes from the disc via a spiral arm, leading to the fragmentation of the disc and the formation of a new protostar. The end state is a hierarchical triple system. Frame 6 shows the binary, which has cleared a gap in the circumbinary material.}
\end{figure*}

\begin{figure*}
\plottwo{\figpath/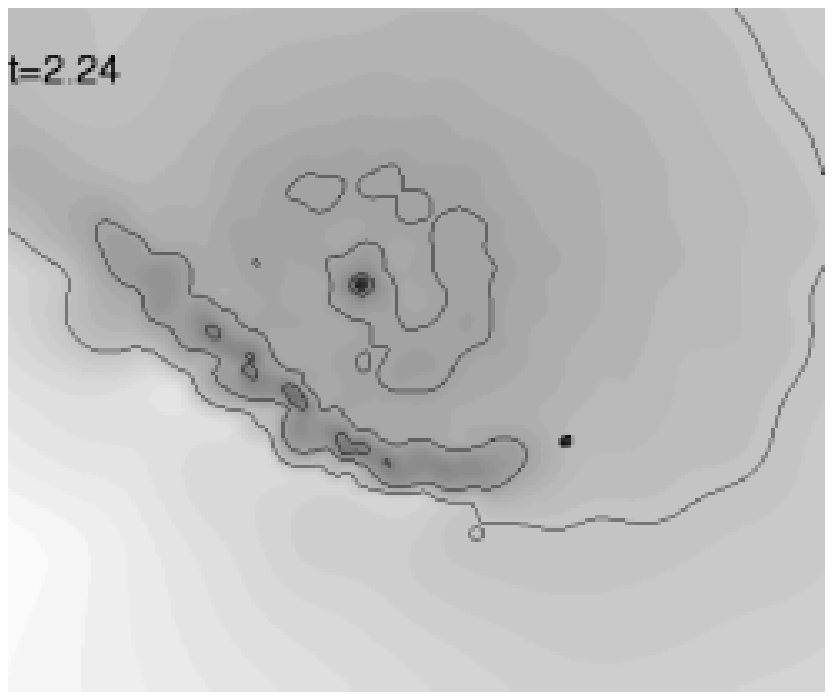}{\figpath/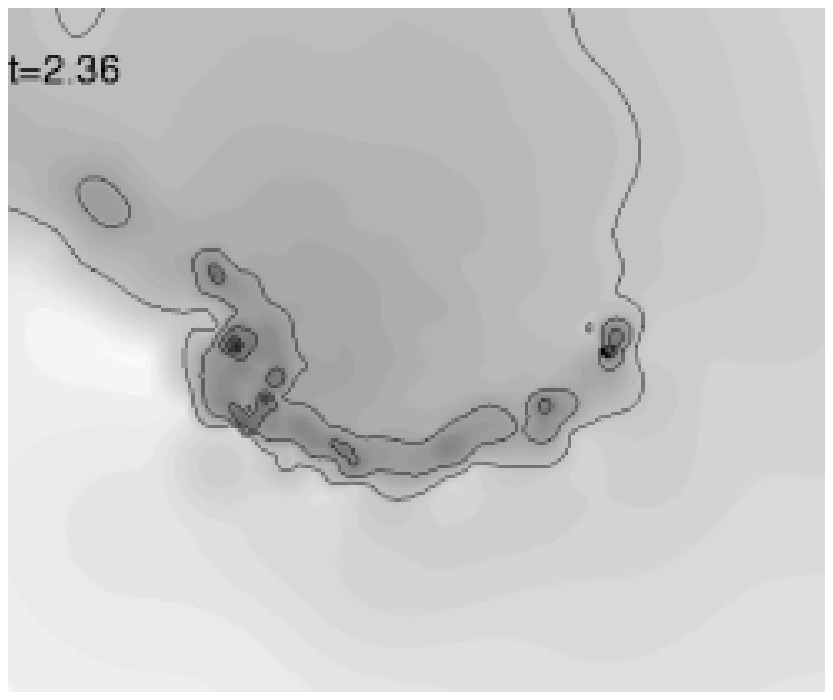}\\
\vspace{6pt}
\plottwo{\figpath/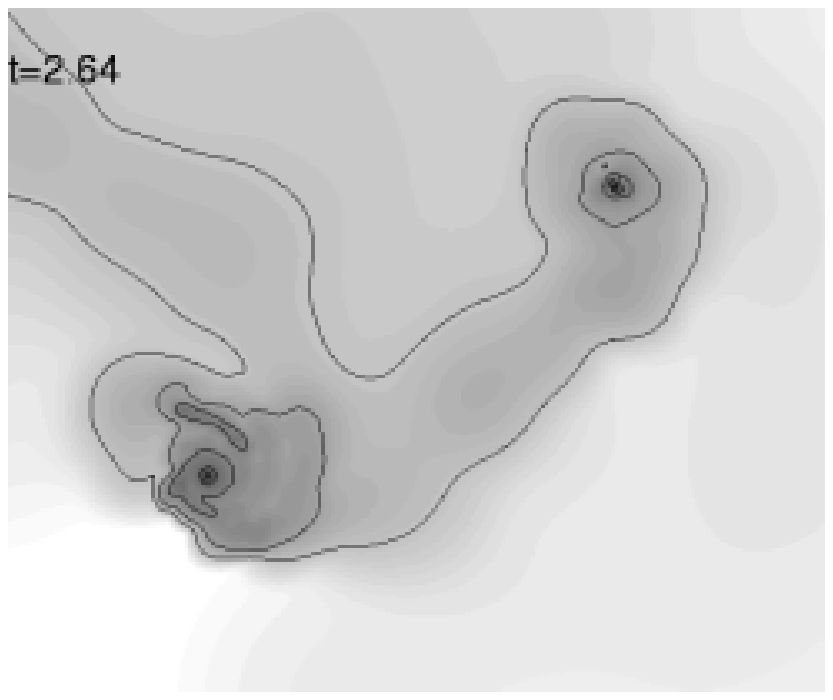}{\figpath/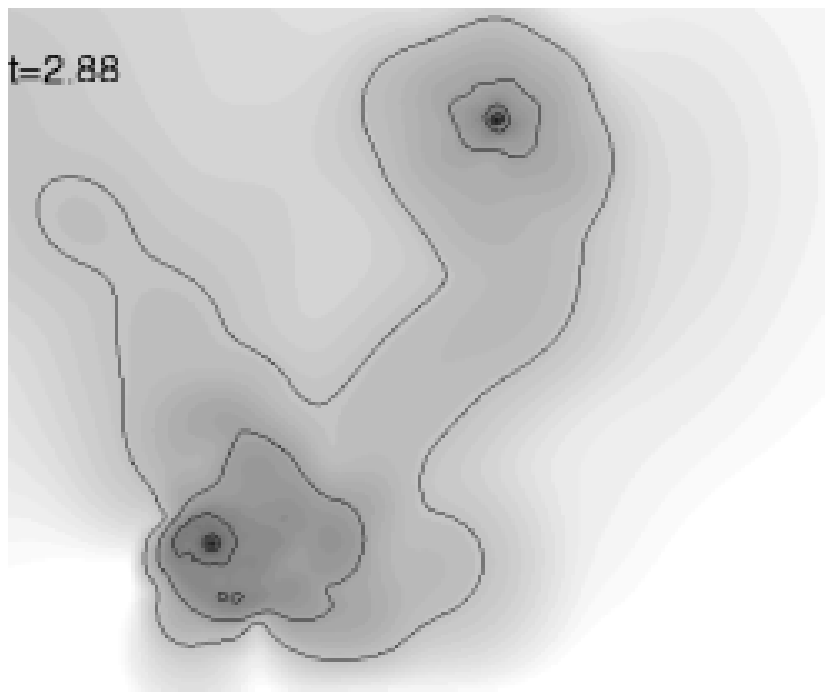}\\
\vspace{6pt}
\plottwo{\figpath/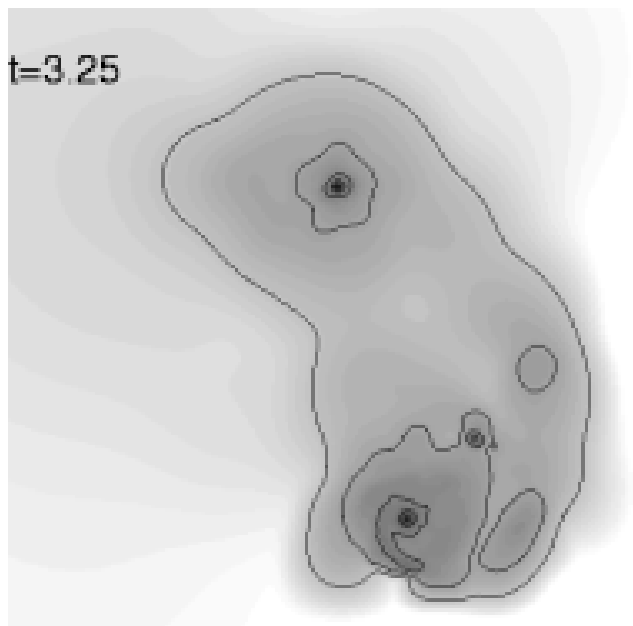}{\figpath/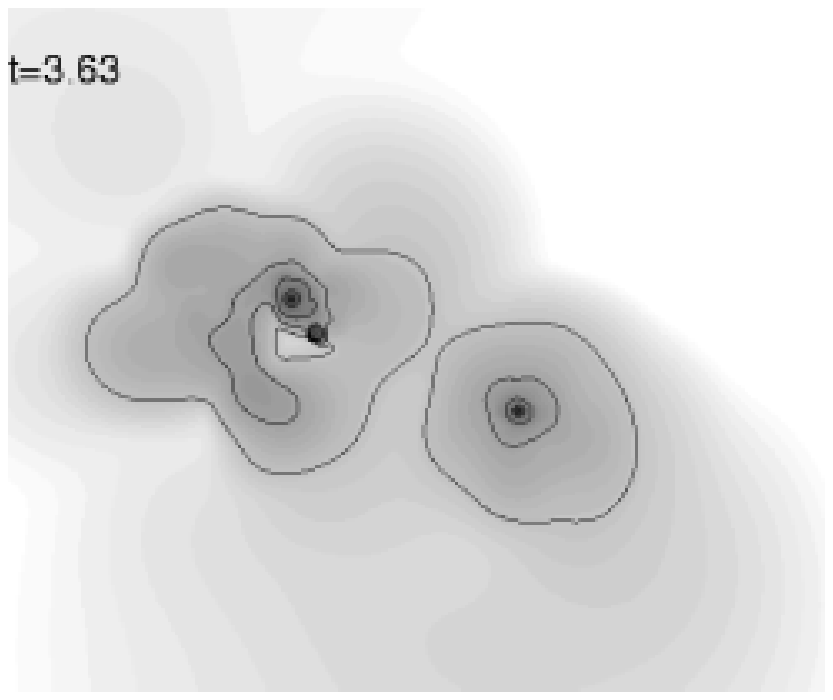}\\
\caption[Simulation ds17 : $\phi=\pi$, $r_{peri}=500 \mbox{AU}$. 2000 $\times$ 1500 AU region.]{\label{fig:ds17}Simulation ds17 : $\phi=\pi$, $r_{peri}=500 \mbox{AU}$. 2000 $\times$ 1500 AU region. Contour levels at [0.3,  3, 30, 300] $\mbox{g cm}^{-2}$. The perturber approaches from the left-hand side. As it passes through the disc, a trailing shock forms behind it. This dissipates energy, and the shocked material begins to fragment and falls onto the primary star. This causes the material around the star to become unstable and it repeatedly throws off spiral arms until it eventually fragments to form a binary companion to the primary star. The end state is a hierarchical triple system.}
\end{figure*}

We have also carried out a number of tests using higher resolution, starting the discs further apart, using different initial surface-density profiles, and omitting to settle the discs in isolation before the start of the encounter. No significant differences in the results are seen. Appendix B presents results obtained when the disc is still modelled with a single layer of particles, but with more particles (11299 instead of 2000) and hence greater resolution. Appendix C presents results obtained when the disc is resolved in the third dimension (parallel to the spin axis) using 13713 particles. In both cases the results are remarkably similar to those obtained with the standard resolution, and this gives us confidence that the basic physics is not being corrupted by lack of resolution.

An additional test simulation has been carried out to investigate the secular stability of an isolated disc, and to determine whether the growth of intrinsic gravitational instabilities within the disc leads to its fragmentation. The evolution of the disc has been followed for 20 orbital periods at the outer edge. One- and two-armed spiral instabilities develop within the disc but act only to redistribute angular momentum within the disc, leading to accretion onto the centre and the spreading of the outer regions of the disc. This is as expected, as the disc that we have chosen to model has a minimum value of the Toomre $Q$-parameter \cite{toomre} of 2.4, and instability to fragmentation is expected only for $Q \stackrel{<}{_{\scriptstyle \sim}} 1$ \cite{laughlin:rozyczka}. The relatively high $Q$ of the disc means that we should not expect the disc to retain the $r^{-3/2}$ surface density profile derived by Lin \& Pringle \shortcite{lin:pringle90}.

\begin{table}
\begin{tabular}{|c|c|c|c|c|} \hline
Run    & $r_{peri}/r_{disc}$ & $\phi$   \\ \hline \hline
ds01-04 &  0.5, 1.0, 1.5, 2.0 & 0 \\ \hline
ds05-08 &  0.5, 1.0, 1.5, 2.0 & $\pi/4$ \\ \hline
ds09-12 &  0.5, 1.0, 1.5, 2.0 & $\pi/2$ \\ \hline
ds13-16 &  0.5, 1.0, 1.5, 2.0 & $3\pi/4$ \\ \hline
ds17-20 &  0.5, 1.0, 1.5, 2.0 & $\pi$ \\ \hline
\end{tabular}
\caption[]{\label{runs}
List of simulations}
\end{table}

\section{Simulations}
\label{sect:sims}

The figures presented in this section show the results of two  of the simulations conducted. The figures are grey-scale plots of column-density, in which the shading is logarithmically-scaled. These plots are overlaid with contours of constant surface density, that are equally separated in log-space. Occasionally an extra contour has been added in order to highlight an important feature of the figure, such as a shock or spiral arm. The contour levels used are given in the caption to the figure. The time at each figure, in units of $10^{4}$ years, is also given, and is calculated from the start of the simulation, when the two stars are at a separation of 5$r_{disc}$. The individual frames within a figure, when referred to within the text or a caption, are labelled from top left to bottom right in rows, so that frames 1 and 2 are on the top row, frames 3 and 4 on the next row and so on.

\subsection{run ds02 : prograde coplanar encounter}

Encounters with $\phi=0$ are prograde, coplanar encounters. The results for run ds02 are shown in Fig.~\ref{fig:ds02}. The periastron distance is 1000AU. In this simulation the perturber just grazes the edge of the disc (frame 1). Disc material flows from the primary onto the secondary via a spiral arm or bridge (frame 2), while a second, opposing arm is formed in the disc (frame 3). As the two stars move away, the accretion arm between them breaks off, but the disc around the primary star remains highly perturbed (frame 4). This $m=2$ perturbation leads to the fragmentation of the disc around the primary, and the formation of a binary companion to the primary (frame 5). This newly-formed binary system is surrounded by some circumbinary material in which it soon clears a gap (frame 6). By the end of the encounter, the perturbing star has accreted a disc of radius 475AU and mass 0.08$\mbox{M}_{\sun}$. The primary star has a small circumstellar disc, of radius 60AU and mass 0.1$\mbox{M}_{\sun}$. Its binary companion has mass 0.04$\mbox{M}_{\sun}$ and no circumstellar material. The two are in an orbit with eccentricity 0.1 and periastron 90AU. They are surrounded by a circumbinary disc of mass 0.15$\mbox{M}_{\sun}$ and radius 350AU. The binary system and the perturbing star are in an orbit that has a periastron of 970AU and eccentricity 0.89.

\begin{figure*}
\plottwo{\figpath/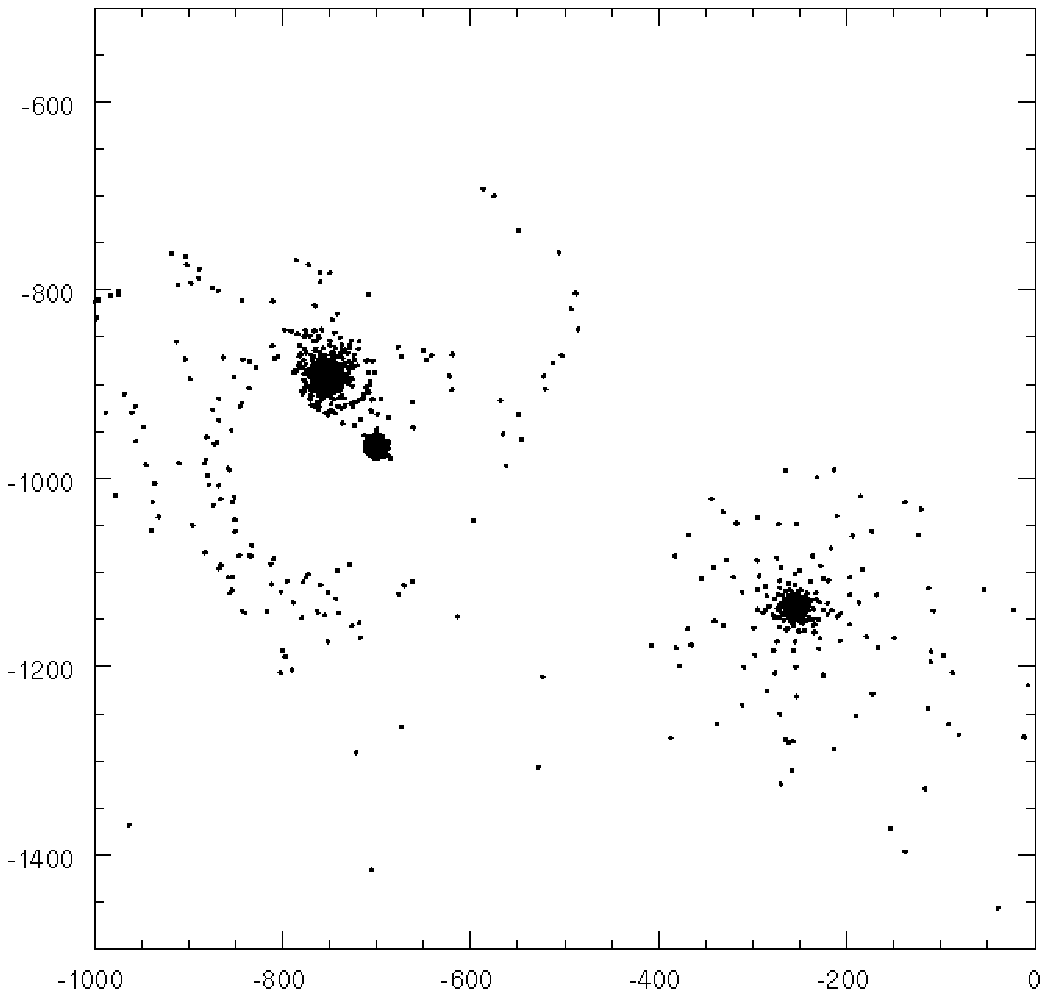}{\figpath/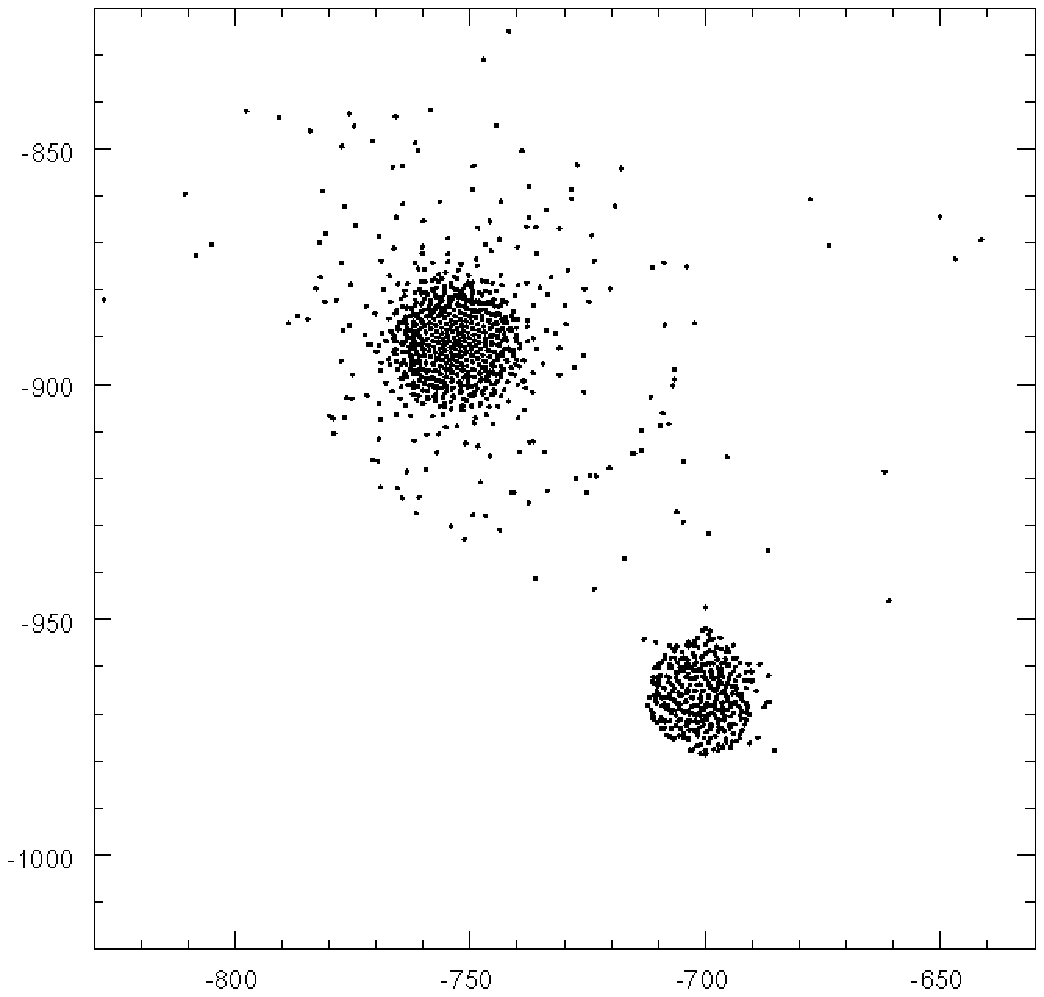}\\
\caption[Simulation ds17 - particle plot]{\label{fig:ds17b}Simulation ds17 : $\phi=\pi$, $r_{peri}=500 \mbox{AU}$. Positions of SPH particles. The left-hand figure shows the triple system at the end of the simulation. The right-hand frame shows a close-up of the primary and its companion. The primary and its disc consist of 1010 particles, whilst its companion consists of 380 particles.}
\end{figure*}

\subsection{run ds17 : retrograde coplanar encounter}

Encounters with $\phi=\pi$ are retrograde coplanar encounters, in which the spins of the disc and the orbit are anti-parallel. The results of run ds17 are shown in Fig.~\ref{fig:ds17}. This encounter has periastron 500AU. The perturber approaches from the left-hand side of the disc and orbits anti-clockwise about it. As the perturber passes through the disc, a trailing shock forms (frame 1), and the material in the shock begins to fragment. Because of the exchange of momentum between the perturber and the shocked material, the shocked material (including the fragments) is no longer centrifugally supported relative to the primary star, and it falls into the region around the primary star (frame 2). This causes the material around the primary star to spin-up and become rotationally unstable. As a consequence, it repeatedly throws off spiral arms, until eventually one of the arms becomes self-gravitating and condenses out to produce a binary companion to the primary star. By the end of the simulation, a large fraction of the original extended circumstellar disc around the primary star has been converted into a much more compact circumstellar disc, having mass 0.26$\mbox{M}_{\sun}$ and radius 50AU. The primary star's binary companion has mass 0.10$\mbox{M}_{\sun}$, and the two are on an orbit with eccentricity 0.07 and periastron 90AU. The perturbing star has accreted a disc of mass 0.08$\mbox{M}_{\sun}$ and radius 200AU. It is bound to the binary, with an eccentricity of 0.50 and a periastron of 320AU. Figure \ref{fig:ds17b} shows particle plots of the triple system at the end of the simulation, and of the primary star and its binary companion. Note that the compact discs shown here are not resolved. Because of gravity softening and shear viscosity, the particles have formed a pseudo-lattice.

\begin{figure*}
\plottwo{\figpath/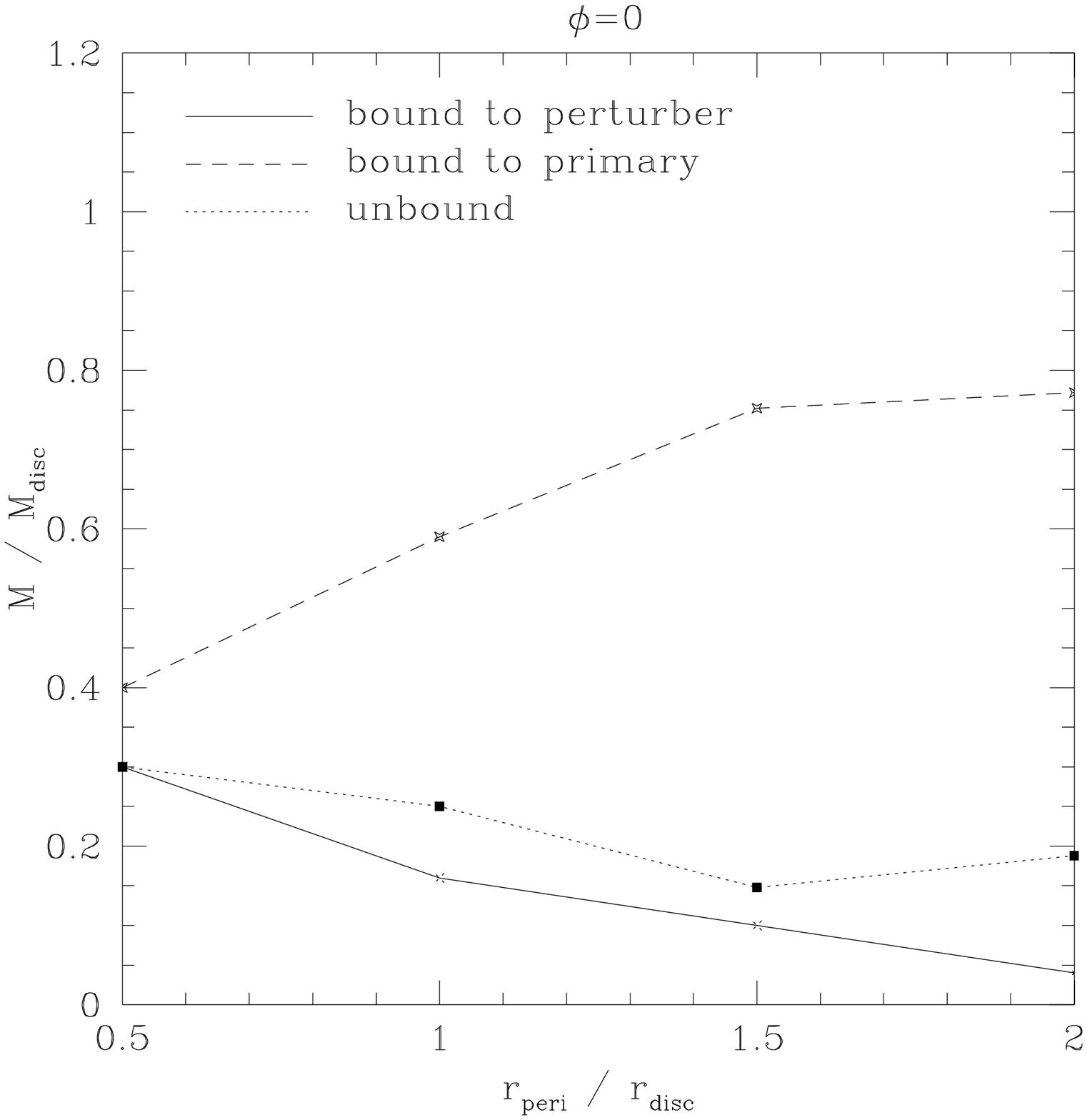}{\figpath/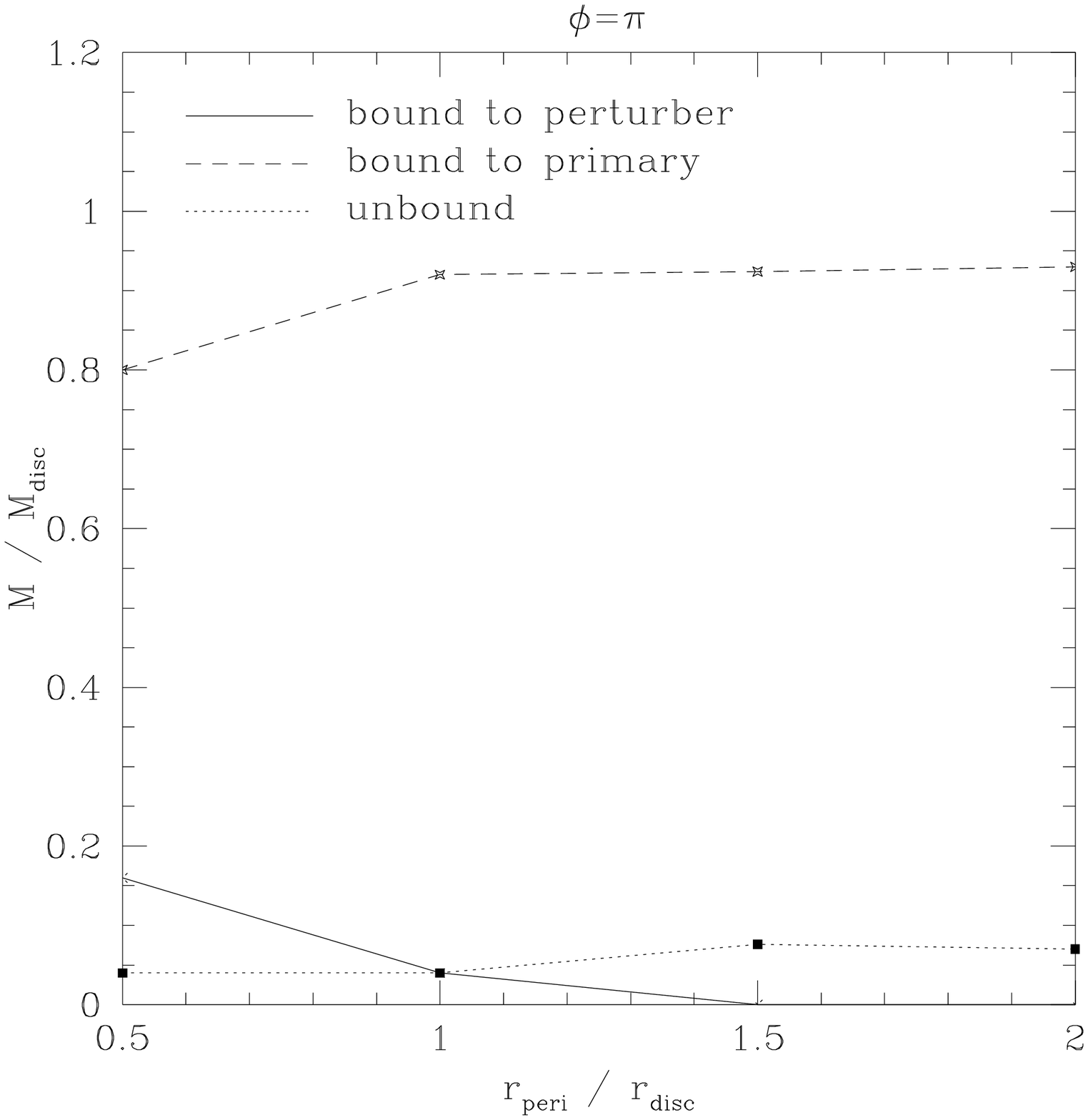}\\
\caption[Mass transfer for coplanar disc-star encounters]{\label{fig:cpsmass}The fraction of the disc mass that ends up bound to the primary star, bound to the perturber and unbound, for coplanar prograde ($\phi=0$) and coplanar retrograde ($\phi=\pi$) encounters.}
\end{figure*}

\begin{figure*}
\plottwo{\figpath/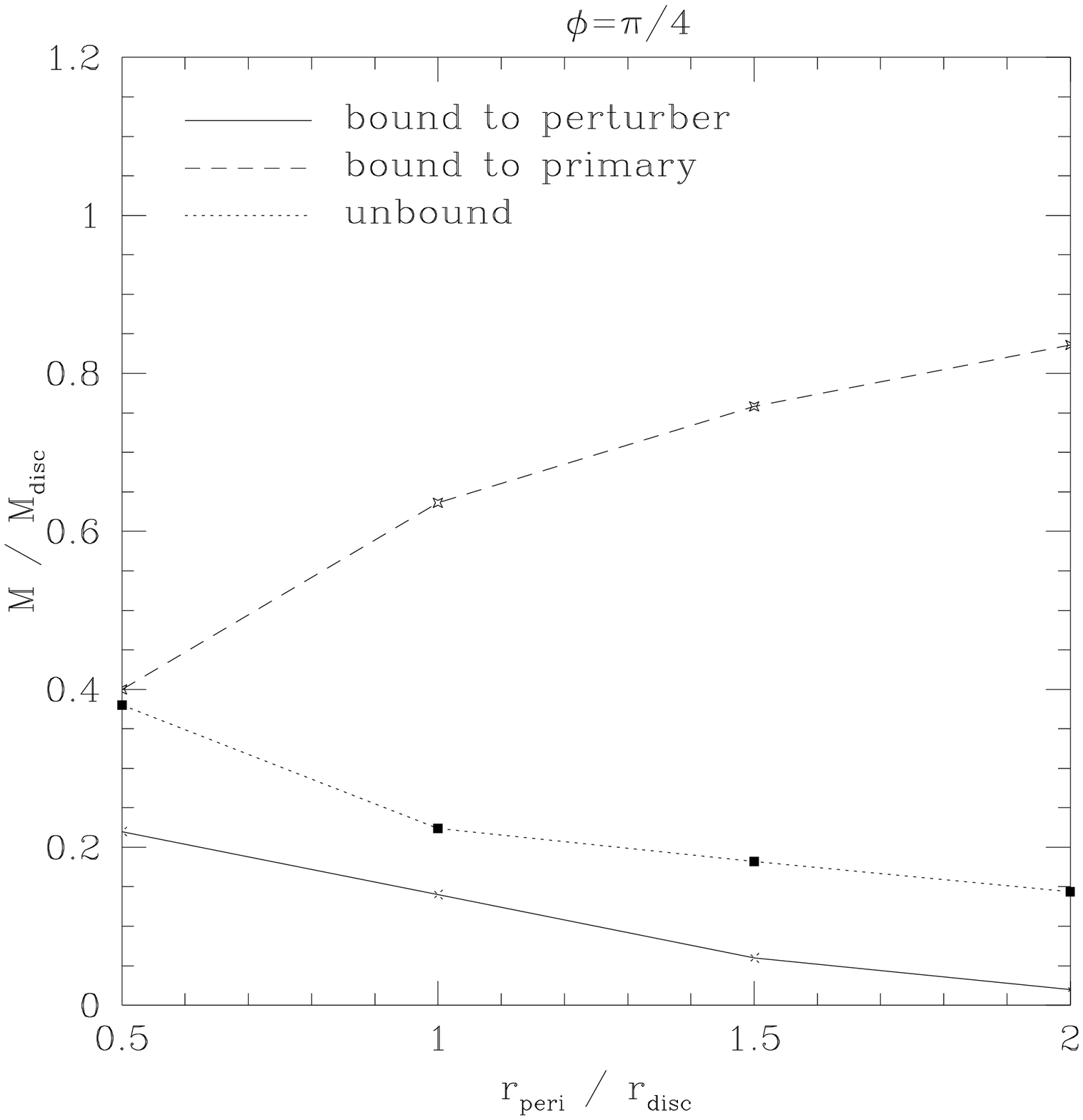}{\figpath/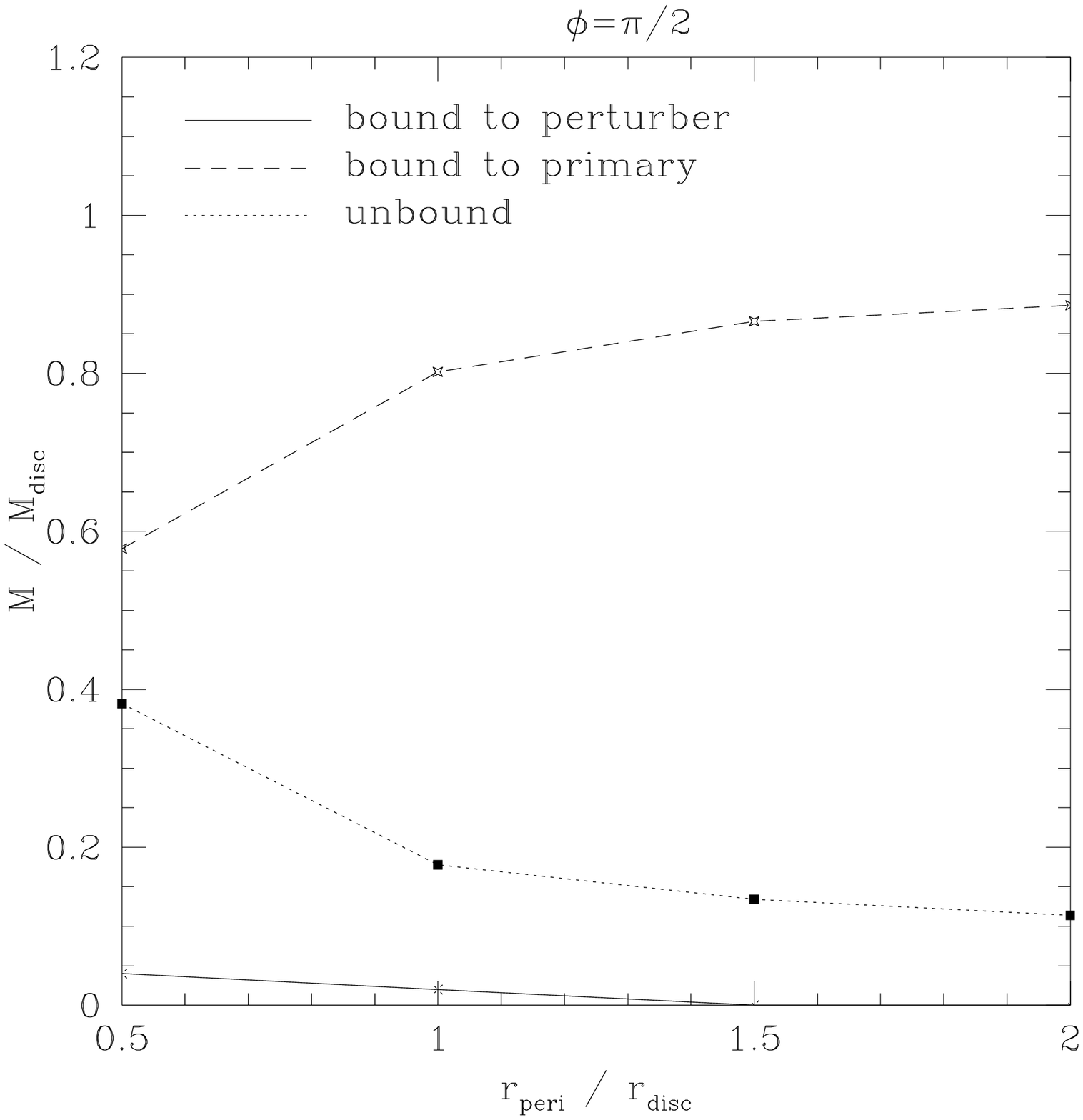}\\
\vspace{6pt}
\plottwo{\figpath/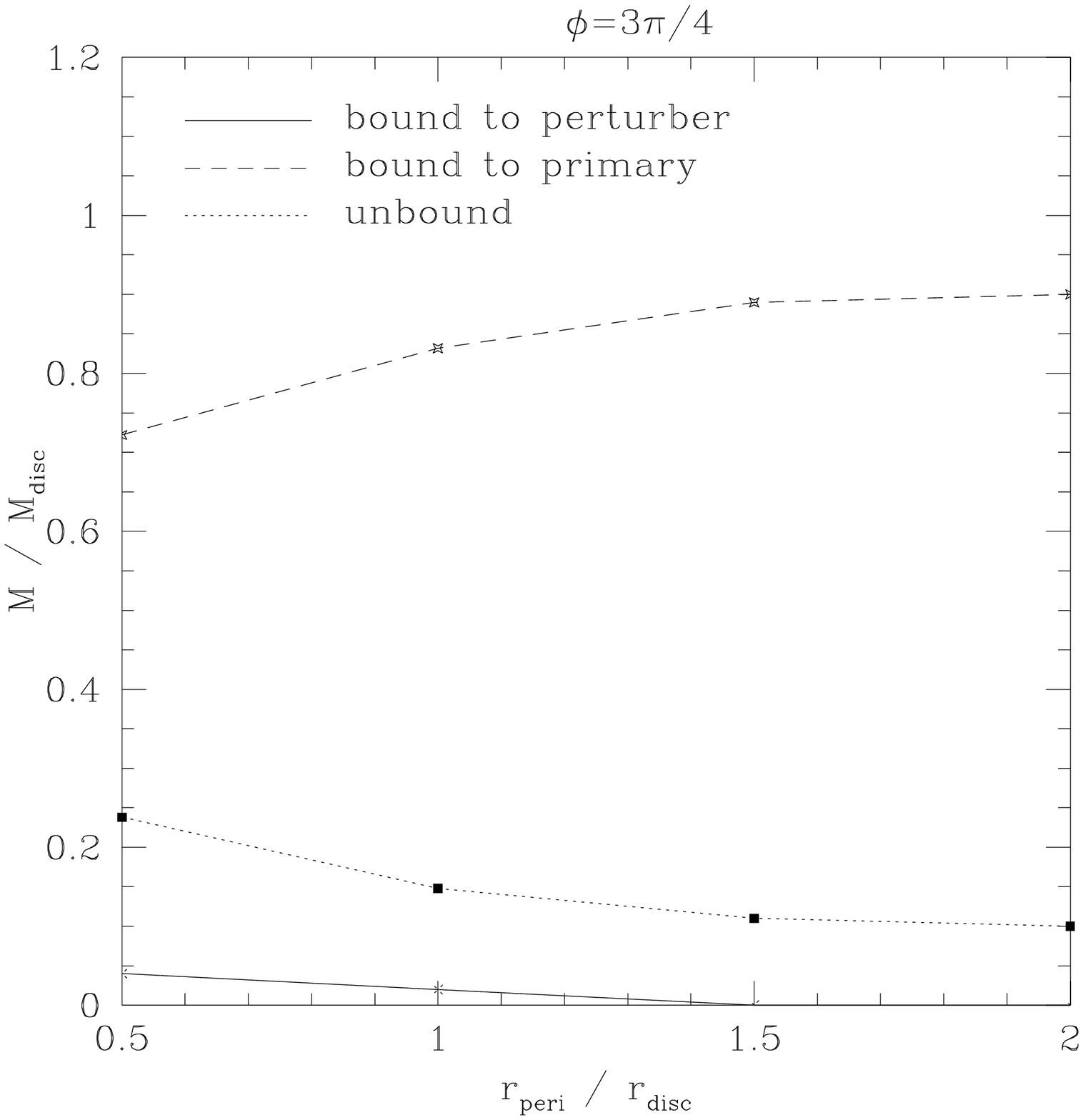}{\figpath/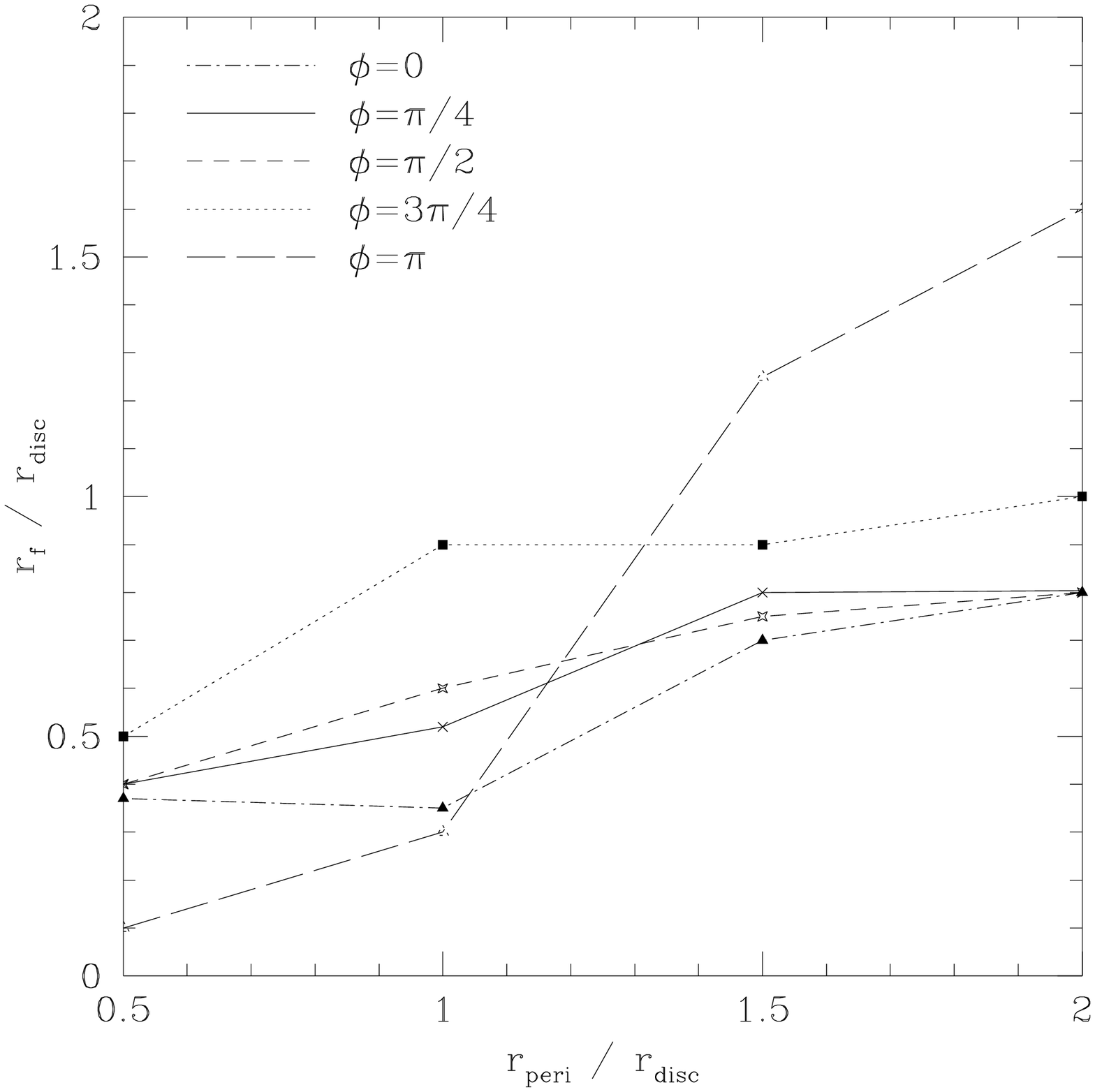}\\
\caption[Mass transfer and disc sizes for disc-star encounters]{\label{fig:ncpsmass}Frames 1-3 show the fraction of the disc mass that ends up bound to the primary star, bound to the perturber and unbound, for non-coplanar encounters with $\phi=\pi/4$, $\pi/2$ and $3\pi/4$ respectively. Frame four shows the size of the disc remaining around the primary star after the encounter for all simulations.}
\end{figure*}

\begin{figure*}
\plottwo{\figpath/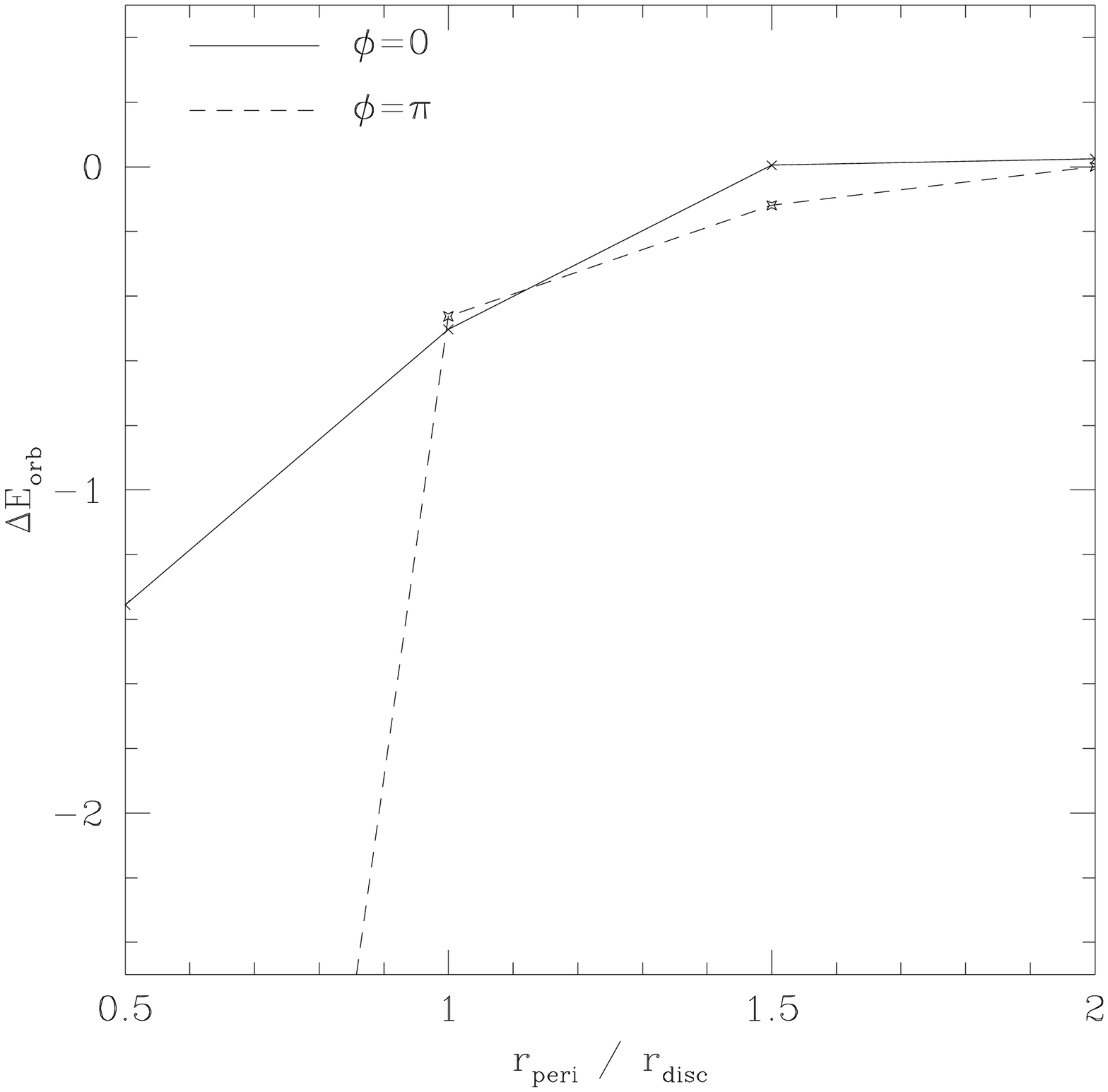}{\figpath/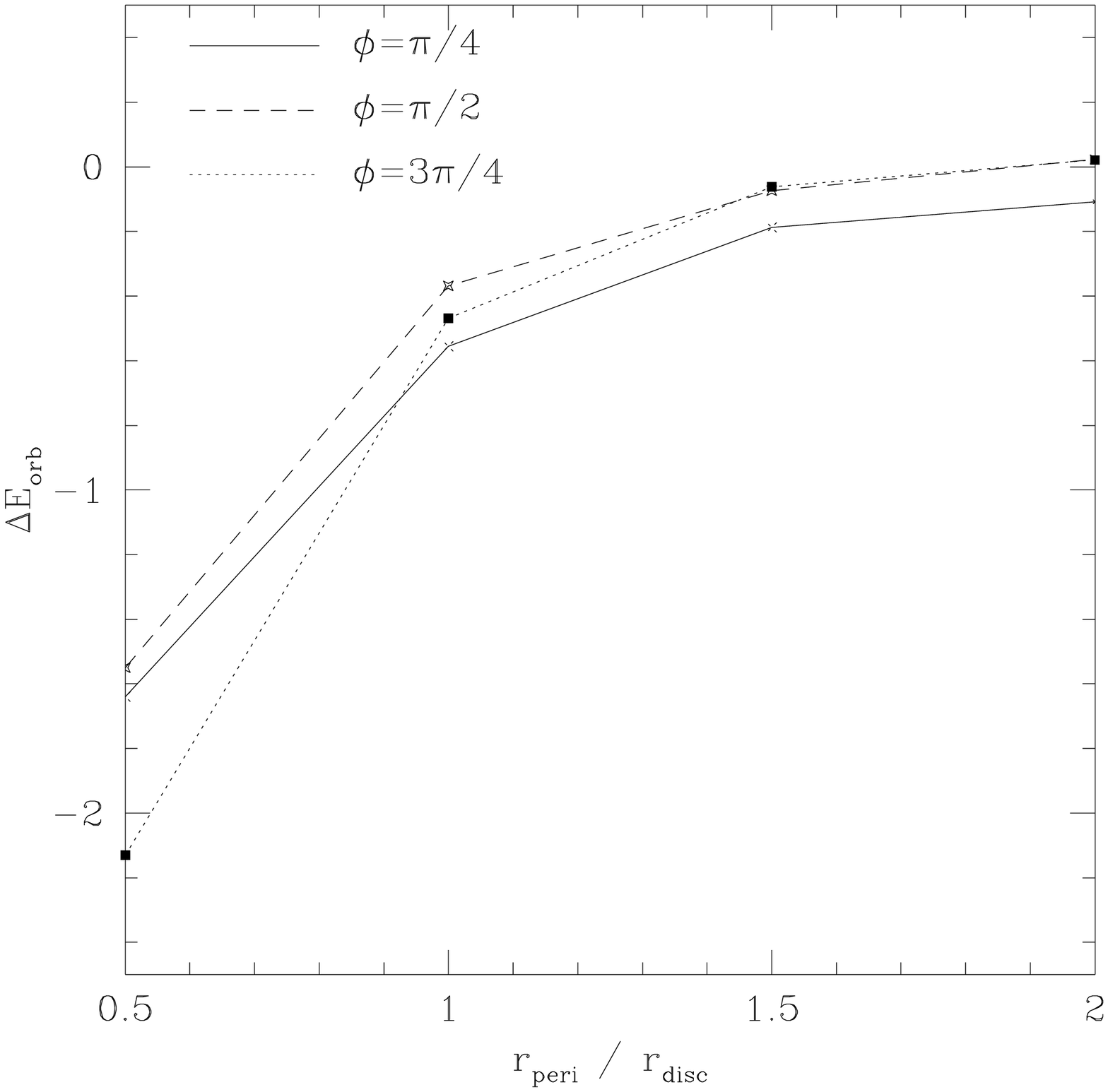}\\
\caption[Energy transfer for disc-star encounters]{\label{fig:staren}Energy transfer to the binary orbit, for coplanar (left-hand figure) and non-coplanar (right-hand figure) encounters.}
\end{figure*}

\section{Results}
\label{sect:res}

The energy of an orbit is given by 
\begin{equation}
E=-\frac{G M_{1} M_{2} (1-e)}{2r_{peri}},
\end{equation}
where $M_{1}$ and $M_{2}$ are the masses of the two bodies, and $e$ and $r_{peri}$ the eccentricity and periastron of the orbit respectively. The simulations presented here have initially parabolic orbits, i.e. eccentricity $e=1$ and energy $E=0$. A reduction in the eccentricity of the orbit corresponds to the two stars becoming bound to each other, whilst an increase in the eccentricity means that the two stars have become unbound from each other.

Non-coplanar encounters leave the disc tilted with respect to the orbit, in agreement with previous results (Heller 1993\nocite{heller}, Larwood et al.\ 1996). Penetrating encounters can twist the disc through up to $30^{\circ}$, while even for $r_{peri}=2r_{disc}$, the outer regions of the disc end up tilted by several degrees. Such a tilt can have a large effect upon the observed spectrum of the disc, due to the reprocessing of light from the central star (Terquem \& Bertout 1993\nocite{terquem:bertout93}, 1996\nocite{terquem:bertout96}).

\subsection{Disc truncation}

The prograde coplanar encounters (ds01-ds04) should be most damaging to the primary disc (cf. Toomre \& Toomre 1972\nocite{toomre:toomre}), and one might expect them also to have the largest effect on the orbit. Conversely one might espect retrograde coplanar encounters to have the least effect. Clarke \& Pringle (1993\nocite{clarke:pringle93}, hereafter CP93) find that truncation of the primary disc is indeed greatest for prograde coplanar encounters. However, Hall, Clarke \& Pringle (1996\nocite{hall}, hereafter HCP96) find that it is retrograde coplanar encounters which remove most energy from the binary orbit. For prograde coplanar encounters with $r_{peri} > r_{disc}$, a corotation resonance can actually lead to the transfer of energy {\it to} the binary orbit, causing the stars to become unbound from each other.

Figure \ref{fig:cpsmass} shows the fraction of the disc mass that ends up bound to the primary star, bound to the perturber, and unbound, for the coplanar encounters we have simulated, whilst the first three frames of Fig.~\ref{fig:ncpsmass} show the corresponding figures for the non-coplanar encounters. Frame 4 of Fig.~\ref{fig:ncpsmass} shows the size of the disc remaining around the primary star. Where the encounter has resulted in the formation of a binary surrounded by a circumbinary disc, the size of the circumbinary disc is used.

Our results are, for the most part, in good general agreement with those of CP93, with the prograde coplanar encounters being the most destructive to the disc, and the retrograde coplanar encounters the least destructive. It can be seen from Fig.~\ref{fig:cpsmass} that the coplanar prograde encounters always lead to more mass being captured by the perturber, and more mass becoming unbound from the system -- than the corresponding coplanar retrograde encounters. For encounters with a prograde element and $r_{peri}>r_{disc}$, i.e. periastron greater than 1000AU, there is a marked reduction in the disruptiveness of the encounters, so that for these non-penetrating encounters, those with $\phi=\pi/4$ actually truncate the disc less than the corresponding orthogonal encounters. This can be attributed to the corotation resonance identified by HCP96.

The two simulations that are not in agreement with CP93 are ds17 and ds18, which are retrograde coplanar encounters with periastron 500 and 1000AU respectively, i.e.\ $r_{peri}=0.5r_{disc}$ and $r_{peri}=r_{disc}$. These encounters are the most destructive of all, in terms of the size of the disc remaining after the encounter. The reason that these encounters are so destructive is that as the perturber passes through the disc, it generates a trailing shock. This shock dissipates enormous amounts of energy, and thereby triggers fragmentation of the disc, so that both ds17 and ds18 end up as multiple systems. The code used by CP93 did not model the self-gravity of the disc or the hydrodynamic forces, and so was unable to model the shocks.

\subsection{Energy exchange}

Figure \ref{fig:staren} shows the change in energy of the orbit of the primary and perturber, for both the coplanar and non-coplanar encounters. The energy is plotted in code units, in which the total binding energy of the disc is 6.6 and the binding energy outside of 0.5$r_{disc}$ is 0.82. We find that for non-penetrating ($r_{peri}>r_{disc}$), non-coplanar encounters, those with $\phi=\pi/4$ are the most destructive, with the orthogonal and $\phi=3\pi/4$ encounters having approximately the same effect. For penetrating encounters, those with a retrograde element can dissipate energy via a trailing shock as they pass through the disc, so that when $r_{peri}=0.5r_{disc}$, the encounter with $\phi=3\pi/4$ removes 30\% more energy from the orbit than the corresponding $\phi=\pi/4$ encounter.

For prograde coplanar encounters, we find, in agreement with the results of HCP96, that encounters with periastron outside the disc radius lead to the transfer of energy to the binary orbit, and the subsequent unbinding of the system, whilst the retrograde encounters remove energy from the orbit. For the grazing encounters, with $r_{peri}=r_{disc}$, the prograde encounter removes marginally more energy than the corresponding retrograde encounter. For the penetrating encounters, with $r_{peri}=0.5r_{disc}$, the retrograde encounter is dominated by the energy loss in the trailing shock behind the perturber, and so dissipates more than five times the energy lost in the corresponding prograde encounter.

McDonald \& Clarke \shortcite{mcdonald:clarke} assumed in their calculations that the energy removed from the orbit during an encounter was equal to the binding energy of the disc outside of periastron. From the results in Fig.~\ref{fig:staren}, it can be seen that the change in orbital energy is in fact approximately twice the binding energy at periastron for $r_{peri}=0.5r_{disc}$, for all except the retrograde coplanar encounter, in which much more energy is dissipated via the trailing shock.

\begin{figure}
\plotsmall{\figpath/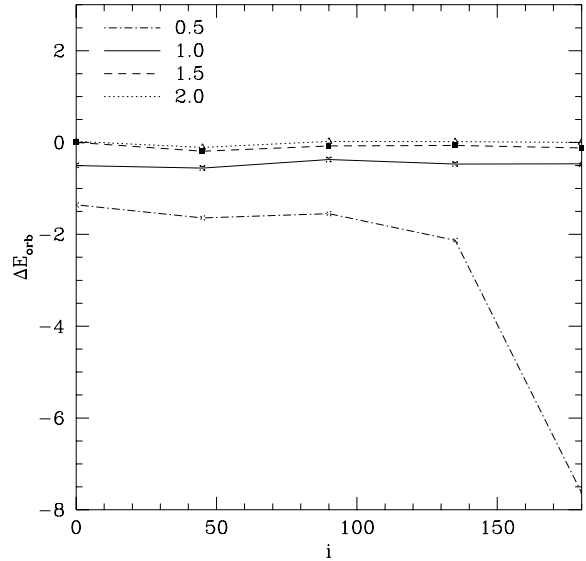}\\
\caption[Energy transfer for disc-star encounters]{\label{fig:inc_en}Energy transfer to the binary orbit, as a function of inclination, for $r_{peri}/r_{disc}$=0.5, 1.0, 1.5, 2.0}
\end{figure}

\begin{figure}
\plotsmall{\figpath/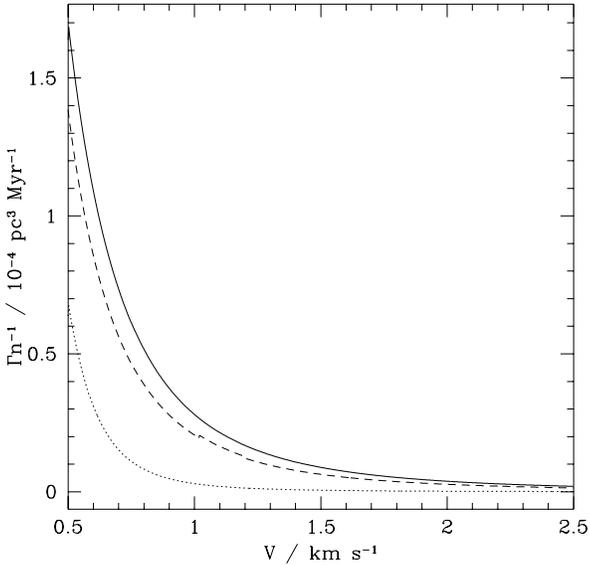}\\
\caption[Capture rate for disc-star encounters]{\label{fig:starcap}Capture rate per unit number density for disc-star encounters, as a function of velocity dispersion. The solid line shows the rate for randomly distributed inclinations, the dashes show the rate assuming all encounters are prograde coplanar, and the dots show the rate corrected for previous disruption of discs.}
\end{figure}

\subsection{Capture rates}

Figure \ref{fig:inc_en} shows the same data as presented in Fig.~\ref{fig:staren}, but now with the change in orbital energy plotted as a function of the inclination of the encounter, for the four different periastra investigated. These data can be used to calculate the cross-section for the formation of binary systems by capture, and hence to determine if the energy dissipation during disc-star encounters provides an effective mechanism for binary formation. In order to do this, we follow the method set out by Heller \shortcite{heller95}.

The cross-section for an interaction with periastron $r$ for two stars each of mass $m$, that have a relative velocity $v$ at large separation, is \cite{binney:tremaine}
\begin{equation}
\sigma(r)=\pi r^{2} + \frac{4 \pi Gmr}{v^{2}}.
\end{equation}
The cross-section for capture during an encounter of inclination $i$ and velocity $v$ is therefore 
\begin{equation}
\Psi(v,i)= \sigma(r_{c}),
\end{equation}
where $r_{c}(v,i)$ is the maximum periastron for an inclination of $i$ for which capture will occur at a velocity $v$, that is the largest periastron for a given $i$ that satisfies
\begin{equation}
\Delta E_{\mbox{{\footnotesize orb}}} < - \frac{1}{2} \mu v^{2},
\end{equation}
where $\mu$ is the reduced mass of the system, and is equal to $m/2$. The value of $r_{c}$ for a given $v$ and $i$ is calculated from the data in Fig.~\ref{fig:inc_en} by using two-dimensional spline interpolation \cite{press92}.

If the alignments of encounters are randomly distributed then the mean cross-section for capture is
\begin{equation}
\sigma_{c} = \frac{1}{2} \int^{\pi}_{0} \Psi(v,i) \sin i \: di.
\end{equation}
If we assume that the stars in a cluster have a Maxwellian velocity distribution, with a one-dimensional velocity dispersion $V$, and that the number density of stars is $n$, then the capture rate per star is given by
\begin{equation}
\Gamma_{c} = \frac{n}{2 \sqrt{\pi} V^{3}} \int^{\infty}_{0} e^{-v^{2}/4V^{2}} \sigma_{c}(v) v^{3} dv.
\end{equation}

The change in $\Gamma_{c}$ caused by extrapolating the data from the inner set of data points at $r_{peri}=0.5r_{disc}$ to $r_{peri}=0$, compared to the case when it is assumed that encounters inside $0.5r_{disc}$ have no effect, is approximately 10\%. It is therefore likely that the effect of these very close encounters is not significant, because of the low frequency with which they occur, and that the results obtained by extrapolating to $r_{peri}=0$ are not greatly in error.

Figure \ref{fig:starcap} shows the capture rate  per unit number density, $\Gamma_{c}$, for the two cases where encounters happen at random inclinations, and where all of the encounters are prograde coplanar encounters. It also shows the rate $\Gamma^{'}_{c}$ assuming that an encounter with $r_{peri} < r_{disc}$ completely dissipates the disc. This modified rate is calculated using $\Gamma^{'}_{c} = \Gamma^{2}_{c}/\Gamma_{h}$ \cite{clarke:pringle91a}, where $\Gamma_{h}$ is the rate at which encounters with $r_{peri}<r_{disc}$ occur, and is given by
\begin{equation}
\Gamma_{h} = \frac{n}{2 \sqrt{\pi} V^{3}} \int^{\infty}_{0} e^{-v^{2}/4V^{2}} \sigma(r_{disc},v) v^{3} dv.
\label{eq:hitrate}
\end{equation}

\begin{table}
\begin{center}
\begin{tabular}{|c|c|c|c|c|} \hline
Region  & $n / \mbox{pc}^{-3}$ & $V / \mbox{km s}^{-1}$ & $\Gamma_{c} / \mbox{Myr}^{-1}$ \\ \hline \hline
Trapezium (centre) & $10^{4}$ & 1.5 &  0.1 \\ \hline
Trapezium & $2 \times 10^{3}$ & 1.5 &  0.02 \\ \hline
Open cluster & $10^{2}$ & 1 &  $3 \times 10^{-3}$ \\ \hline
GMC & 12 & 2 &  $6 \times 10^{-4}$ \\ \hline
\end{tabular}
\end{center}
\caption[Capture rates for star-disc encounters]
{\label{table:dscap}Capture rates for star-disc encounters in different star-forming environments.}
\end{table}

In fact, as can be seen in Fig.~\ref{fig:ncpsmass}, encounters with $r<r_{disc}$ remove only about 50\% of the mass from the disc. The corrected rate therefore  gives a lower limit to the capture rate, as it overestimates the rate of destruction of discs by encounters, while the uncorrected rate takes no account of the destruction, and so provides an upper limit, with the true value lying somewhere in between.

\begin{figure}
\plotsmall{\figpath/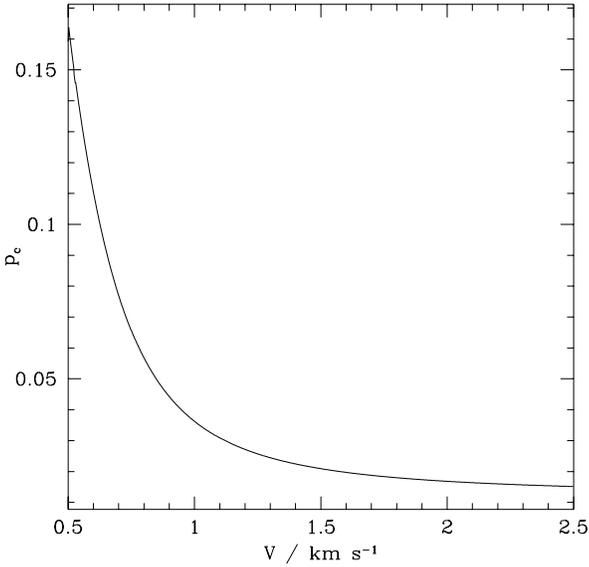}\\
\caption[Probability of capture for disc-star encounter]{\label{fig:sdcap}Mean probability of capture during a disc-star encounter for a randomly-inclined encounter, as a function of velocity dispersion.}
\end{figure}

In order to determine whether capture via disc-star interactions is an important binary formation mechanism, we calculate capture rates in the same stellar environments considered by Clarke \& Pringle \shortcite{clarke:pringle91a}.
 The rates are shown in Table \ref{table:dscap}. Even in the central 0.1pc of the Trapezium cluster, under the most extreme conditions, capture can only give a binary fraction of 10\% after 1Myr. Within the low-density environment of a giant molecular cloud, capture gives a binarity of less than 0.1\% after 1Myr, compared to an observed binary frequency of about 80\% for pre-main-sequence stars in such clouds \cite{mathieu94}. Capture during disc-star interactions is therefore probably not an important binary formation mechanism in typical star-forming regions.

\begin{table}
\begin{center}
\begin{tabular}{|c||c|c|c|c|c|} \hline
$r_{peri}/r_{disc}$  & $\phi=0$ & $\phi=\frac{\pi}{4}$ & $\phi=\frac{\pi}{2}$ & $\phi=\frac{3\pi}{4}$ & $\phi=\pi$  \\ \hline \hline
0.5 & 0 & 0 & 0 & 0 & 1\\ \hline
1.0 & 1 & 2 &  0 & 0 & 2 \\ \hline
1.5 & 1 & 0 &  0 & 0 & 0 \\ \hline
2.0 & 0 & 0 &  0 & 0 & 0 \\ \hline
\end{tabular}
\end{center}
\caption[Companion formation rates for star-disc encounters]
{\label{table:dscomp}Number of companions formed by disc fragmentation}
\end{table}

The rates calculated assume that star formation is distributed uniformly across those regions, and the low capture rates are primarily due to the low number of stars that undergo encounters in such distributed star formation. For instance, in the case of the GMC, only 0.5\% of stars will undergo an encounter in 1Myr \cite{clarke:pringle91a}. If star formation is dynamically-triggered, then the stars formed will be much more tightly clustered than is the average across the whole of the star-forming region. For instance, in simulations of star formation triggered by clump-clump collisions in a GMC (Chapman et al.\ 1992\nocite{chapman92}, Turner et al.\ 1995\nocite{turner}, Whitworth et al.\ 1995\nocite{whitworth}), the majority of the stars that are formed undergo encounters. The star formation in such events is coeval, however, so that the initial encounters that occur are likely to be disc-disc rather than disc-star interactions. It is possible for disc-star encounters to occur if a star has been stripped of its disc by a previous encounter.

If, in such dynamically-triggered star formation, almost all of the stars that are formed undergo encounters, then the probability of capture occuring during an encounter is simply $p_{c}=\Gamma_{c}/\Gamma_{h}$, where $\Gamma_{h}$ is in this case the rate of encounters with $r_{peri}< 2r_{disc}$. Figure \ref{fig:sdcap} shows the probability of capture during an encounter as a function of the velocity dispersion. It can be seen that even if every protostar undergoes one disc-star encounter at speed $\sim$ 1 $\mbox{km s}^{-1}$, this will lead to a binary frequency of only a few percent. Capture during disc-star interactions can not therefore be an important binary formation mechanism, even in dense regions of triggered star formation.

\subsection{Fragmentation rates}

In several of the simulations, an encounter acts to trigger instabilities within the disc that lead to the fragmentation of the disc, and the formation of new companions to the primary star. Table \ref{table:dscomp} lists the number of new stars formed by disc fragmentation. The lower number of fragments for $r_{peri}=0.5r_{disc}$ as compared to $r_{peri}=r_{disc}$ is probably due to two effects, the first being that the closer encounter leaves the disc much more truncated, and so with less mass that can fragment, and the second being that the inner regions of the disc are not fully resolved, as discussed in Section \ref{disc:stab}. We calculate the rate of formation of new protostars, all of which end up as companions to the primary, using a method similar to that above. The data from the simulations are used to calculate the number of new stars formed, $f(r,i)$ after an encounter at inclination $i$ and periastron $r$, so that for instance an encounter that led to the formation of a binary companion to the primary star would have $f=1$. Two-dimensional spline interpolation is used to calculate $f$ for a general $r,i$, subject to the constraint that $f(r,i) \geq 0$ for all $r$ and $i$. We then average over inclinations to get the mean number of companions formed after an encounter at periastron $r$,
\begin{equation}
F(r) = \frac{1}{2} \int^{\pi}_{0} f(r,i) \sin i \: di.
\end{equation}
The rate of encounters within a distance $r$ to $r+dr$ of the star is obtained by differentiating Eq.~(\ref{eq:hitrate}) to give 
\begin{equation}
\Gamma^{'}_{h}(r) = \frac{n}{2 \sqrt{\pi} V^{3}} \int^{\infty}_{0} e^{-v^{2}/4V^{2}} \frac{\partial \sigma(r,v)}{\partial r} v^{3} dv,
\end{equation}
where
\begin{equation}
\sigma(r,v) = \pi r^{2} + \frac{4 \pi Gmr}{v^{2}},
\end{equation}
so that the rate of formation of companions by fragmentation is
\begin{equation}
\Gamma_{f} = \frac{n}{2 \sqrt{\pi} V^{3}} \int^{\infty}_{0} e^{-v^{2}/4V^{2}} \frac{\partial \sigma(r,v)}{\partial r} F(r) v^{3} dr dv.
\end{equation}
We also calculate the rate corrected for the destruction of discs by previous encounters, $\Gamma_{f}^{'}=\Gamma_{f}^{2}/\Gamma_{h}$, where $\Gamma_{h}$ is the rate of encounters with $r_{peri}<r_{disc}$.

\begin{figure}
\plotsmall{\figpath/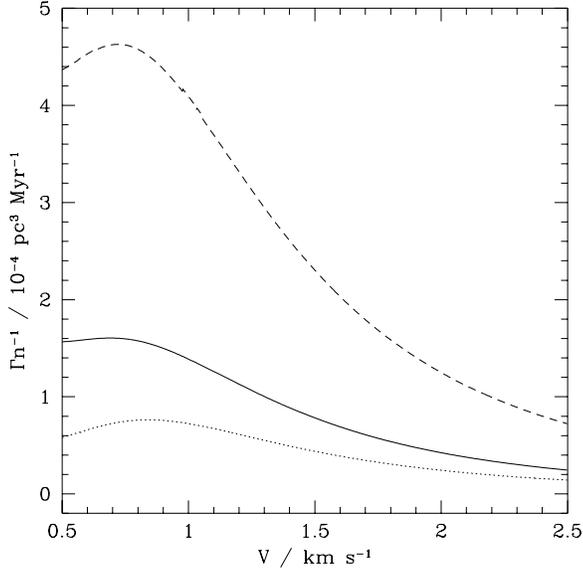}\\
\caption[Companion formation rate for disc-star encounters]{\label{fig:starbin}Rate of formation of companion stars, per unit number density for disc-star encounters, as a function of velocity dispersion. The solid line shows the rate for randomly distributed inclinations, the dashes show the rate assuming all encounters are prograde coplanar, and the dots show the rate corrected for previous disruption of discs.}
\end{figure}

The results are shown in Fig.~\ref{fig:starbin}. In the same way as for the capture calculations, we estimate the rate of formation of companions for different star-forming environments. These results are presented in Table \ref{table:dsbin}. In the central regions of the Trapezium, fragmentation triggered by disc-star interactions can lead to a sizeable proportion of the stars ending up in multiple systems. If all encounters are prograde coplanar, then a star will have on average 1.1 companions after 1Myr. It is important to note, however, that in a region like the Trapezium, the evolution of the whole system will be dominated by the presence of a few massive OB stars. The intense radiation from these OB stars is likely to strip the discs from other stars in the vicinity on a very short time-scale.

For lower-density regions such as a standard GMC, the rate of formation of binary systems by this mechanism is lower, due to the smaller rate of encounters. In the centre of an open cluster, random and prograde encounters can lead to 10\% and 40\% of stars having companions respectively after 10Myr, which is believed to be an upper limit on the lifetime of protostellar discs, with lifetimes an order of magnitude smaller being more likely \cite{strom89a}. The corresponding figures for a GMC are 0.5\% and 1\%. Thus in the case of an open cluster, fragmentation triggered by encounters may be an important binary formation mechanism, but it cannot be a dominant one, while for GMCs it leads to a negligible binary frequency.

\begin{table*}
\begin{center}
\begin{tabular}{|c|c|c|c|c|} \hline
Region  & $n / \mbox{pc}^{-3}$ & $V / \mbox{km s}^{-1}$ & $\Gamma_{f} / \mbox{Myr}^{-1}$ & $\Gamma_{p} / \mbox{Myr}^{-1}$ \\ \hline \hline
Trapezium (centre) & $10^{4}$ & 1.5 & 0.8 & 2.2 \\ \hline
Trapezium & $2 \times 10^{3}$ & 1.5 &  0.2 & 0.4 \\ \hline
Open cluster & $10^{2}$ & 1 &  0.01 & 0.04 \\ \hline
GMC & 12 & 2 &  $5 \times 10^{-4}$ & $1 \times 10^{-3}$ \\ \hline
\end{tabular}
\end{center}
\caption[Companion formation rates for star-disc encounters]
{\label{table:dsbin}Rate of formation of companions for star-disc encounters in different star-forming environments. $\Gamma_{f}$ is the uncorrected rate assuming random inclinations for encounters, $\Gamma_{p}$ is the corresponding rate if all encounters are coplanar prograde.}
\end{table*}

\begin{figure}
\plotsmall{\figpath/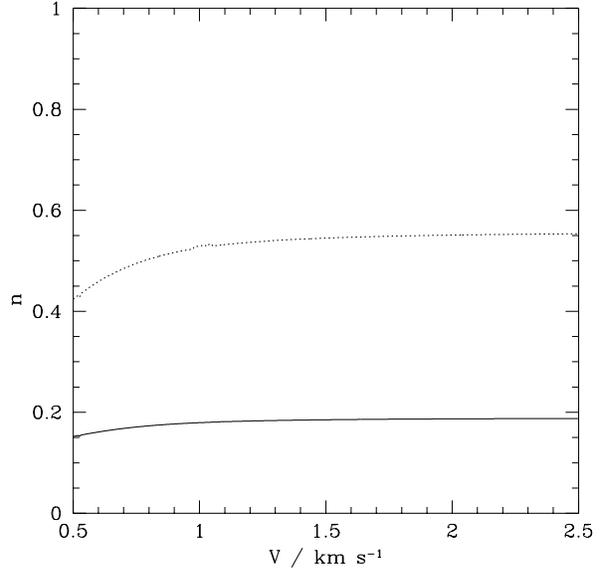}\\
\caption[Mean companion formation rate for a disc-star encounter]{\label{fig:sdnfrag}Mean number of companions formed in a disc-star encounter, as a function of velocity dispersion. The solid line shows the number for randomly distributed inclinations, the dots show the number assuming all encounters are prograde coplanar.}
\end{figure}

Again it is necessary to note that these rates are valid only for distributed star formation. In dynamically-triggered star formation it is possible for nearly every protostar formed to undergo at least one encounter. The mean number of companions formed during an encounter is given by $n=\Gamma_{f}/\Gamma_{h}$ for randomly-oriented encounters, where $\Gamma_{h}$ is the rate of encounters with $r_{peri}<2r_{disc}$, and $n=\Gamma_{p}/\Gamma_{h}$ if all encounters are coplanar prograde. These are shown in Fig.~\ref{fig:sdnfrag}. If there is no preferred inclination for the encounter then the mean number of companions after an encounter is approximately 0.2, whilst if all protostars undergo coplanar prograde encounters, the mean number of companions is 0.5. These correspond to binary fractions of 20\% and 50\% respectively, as compared to an observed pre-main-sequence binary fraction of approximately 50\% \cite{reipurth}. Thus if all protostars undergo one coplanar prograde disc-star encounter, the resulting binary formation by disc fragmentation is consistent with the observed binary fraction. However, the stars formed during dynamically-triggered star formation will be coeval, so that both interacting stars are likely to possess discs. In papers II and III, we present the results of simulations of such disc-disc interactions.

\section{Conclusions}

We have carried out a series of simulations of encounters between protostellar discs and passing stars. The encounters can lead to the removal of energy from the orbit of the encounter, and can cause fragmentation of the disc either via spiral-arm instabilities triggered by the encounter, or, in the case of retrograde coplanar encounters, through the formation of a trailing shock as the star passes through the disc. However, we find that the energy removed from the orbit is not enough to lead to the formation of binary systems by capture at a significant rate. In contrast, disc fragmentation is quite effective in creating new companions for the stars at the centres of discs, provided the density is high enough for frequent star-disc interactions, as for instance in an open cluster environment.

When star formation is dynamically triggered, for instance by the collision of two clumps within a Giant Molecular Cloud, or within a shell swept up by an expanding HII region or stellar wind bubble, interactions are much more likely to occur, and so triggered fragmentation may lead to a much higher binary frequency, particularly if most encounters are prograde coplanar. Dynamically-triggered star formation is likely to be coeval, however, so that during these interactions both protostars will possess discs. We investigate the dynamics of such disc-disc interactions in two subsequent papers.

\section*{acknowledgements}

SJW would like to thank Cathie Clarke for helpful discussions. SJW and ASB acknowledge the receipt of University of Wales studentships. NF acknowledges the receipt of a PPARC studentship. HB and SJW acknowledge the support of a PPARC post-doctoral grant GR/K94157. This work was supported by grants GR/K94140 and GR/L29996 from PPARC.

\appendix

\section{Code tests}

Detailed tests of the numerical method used in this paper were published in Watkins et al.\ \shortcite{watkins}. Among these was the Riemann shock-tube \cite{sod}, which studied the ability of the method to model bulk viscosity correctly. The results of this test were compared to those of an identical test carried out by Turner et al. \shortcite{turner} using the standard artificial viscosity of Monaghan \shortcite{monaghan89}. In the implementation of the test (in both papers) the change in internal energy, $\frac{du}{dt}$, was neglected. In most simulations (including those presented in this paper), a barytropic equation of state is used, with the internal energy being solely a function of density, so that it is not necessary to calculate $\frac{du}{dt}$. However, for the shock-tube, the equation of state is not barytropic, and neglecting this term leads to the entropy remaining constant across the shock, whereas in fact entropy should be generated in the shock. Therefore we have repeated the shock-tube simulation including the $\frac{du}{dt}$ term.

The initial conditions and method used are described in Watkins et al.\ \shortcite{watkins}. When using the artificial viscosity, the rate of change of internal energy, due to pressure and viscous forces, is given by \cite{monaghan92}
\begin{equation}
\frac{du_{i}}{dt}  =  \sum_{j} m_{j} \left[ \frac{P_{i}}{\rho_{i}^{2}} + \frac{P_{j}}{\rho_{j}^{2}} + \frac{1}{2} \Pi_{ij} \right] ({\bf v}_{i} - {\bf v}_{j}) \cdot {\bf \nabla}W_{ij},
\label{inten1}
\end{equation}
where
\begin{equation}
\Pi_{ij} = \left\{ 
\begin{array}{ll} 
\frac{- \alpha \overline{c_{ij}} \mu_{ij} + \beta \mu_{ij}^{2}}{\overline{\rho_{ij}}} & {\bf v}_{ij} \cdot {\bf r}_{ij} < 0; \\
0 & {\bf v}_{ij} \cdot {\bf r}_{ij} > 0;
\end{array}
\right.
\end{equation}
and
\begin{equation}
\mu_{ij} = \frac{\bar{h} {\bf v}_{ij} \cdot {\bf r}_{ij}}{{\bf r}^{2}_{ij} + 0.01 h^{2}},
\end{equation}
where $\overline{c_{ij}}$ is the mean value of the sound speed at particles $i$ and $j$, and $h$ is the SPH smoothing length.

For a Navier-Stokes kinematic bulk viscosity, $\nu$, the viscous contribution to the rate of change of internal energy is given by
\begin{equation}
\frac{du_{i}}{dt} = \nu_{i} ({\bf \nabla} \cdot {\bf v})_{i}^{2}.
\end{equation}
We find that more accurate results are obtained by using a mean of the contributions from particles $i$ and $j$, in a manner similar to that used for the artificial viscosity \cite{hernquist:katz}, so that the contribution becomes
\begin{equation}
\frac{du_{i}}{dt} = \sum_{j} m_{j}  (q_{i} + q_{j}) ({\bf v}_{i}-{\bf v}_{j}) \cdot {\bf \nabla}W_{ij},
\label{inten2}
\end{equation}
where
\begin{equation}
q_{i}  =  \frac{\nu_{i}}{2} \frac{({\bf \nabla} \cdot {\bf v})_{i}}{\rho_{i}}.
\end{equation}
We use the $\beta$-term from the artificial viscosity, with $\beta=1$, and  $\nu_{i}=c_{i} h_{i} / 8$, so that the bulk viscosity should be equivalent to that of the artificial viscosity with $\alpha \sim 1$ \cite{murray}. The total change of internal energy in this simulation is therefore given by taking the sum of Eqs. \ref{inten1} and \ref{inten2}, with $\alpha=0$ and $\beta=1$.

\begin{figure*}
\plotone{\figpath/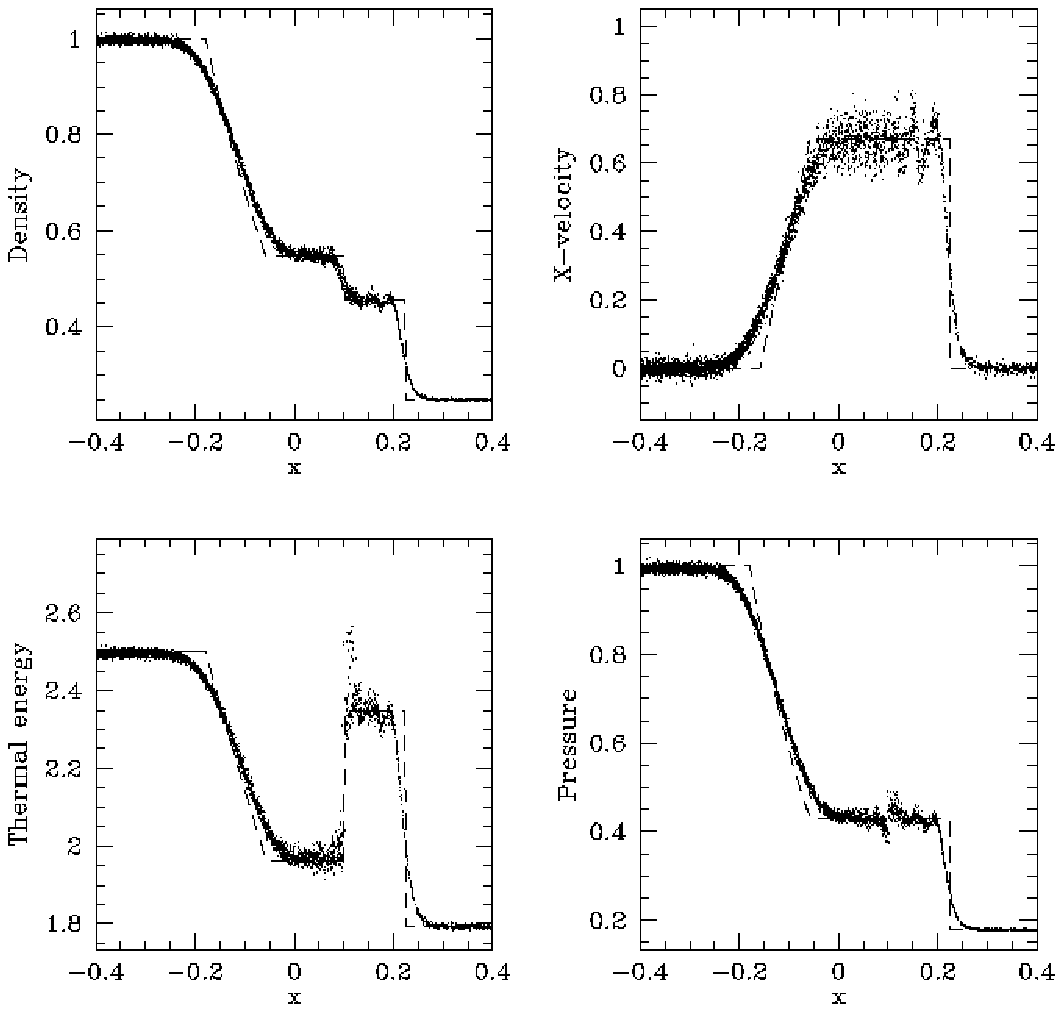}
\caption[]{\label{fig:shock}Results for the Riemann shock-tube at time t=0.16}
\end{figure*}

The results of the test are shown in Fig.~\ref{fig:shock}. The rarefaction wave, contact discontinuity and shock can all be seen. The analytic solution is plotted as a dashed line. The introduction of the $\frac{du}{dt}$ term leads, to more scatter in the velocity in the post-shock region, but the scatter is about the correct mean, and the shock and contact discontinuity are adequately modelled. In particular, the density changes are correct.

When comparing these results with those from other codes, it is important to note (i) that they were obtained with a 3-D code (much more impressive -- but essentially irrelevant -- results obtained with 1-D codes are often presented in this context); and (ii) that the viscosity used here has been fine-tuned to reproduce problems involving shear-flows rather than shocks. Although the results do not fit the analytic solution perfectly, they compare favourably with those obtained by Rasio \& Shapiro 
\shortcite{ras:sha} (their Figure 2) using a comparable number of particles 
($\sim 10^4$).

\section{High resolution run}

\begin{figure*}
\plottwo{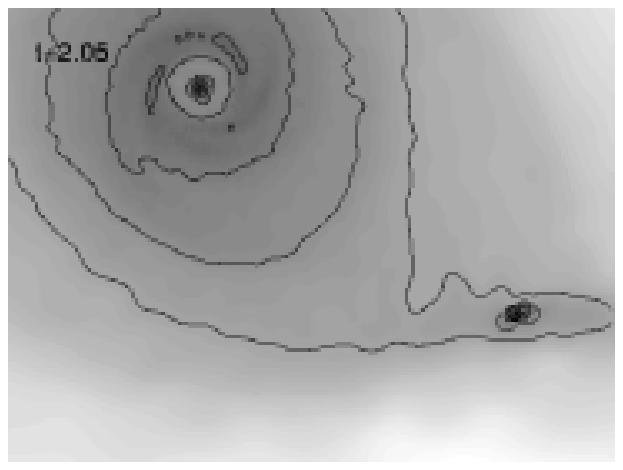}{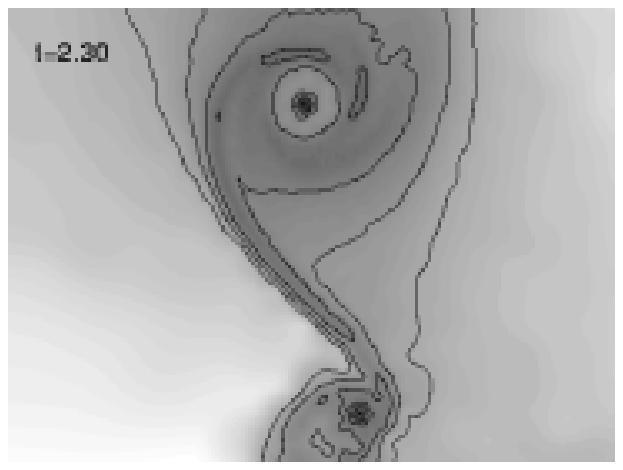}\\
\vspace{6pt}
\plottwo{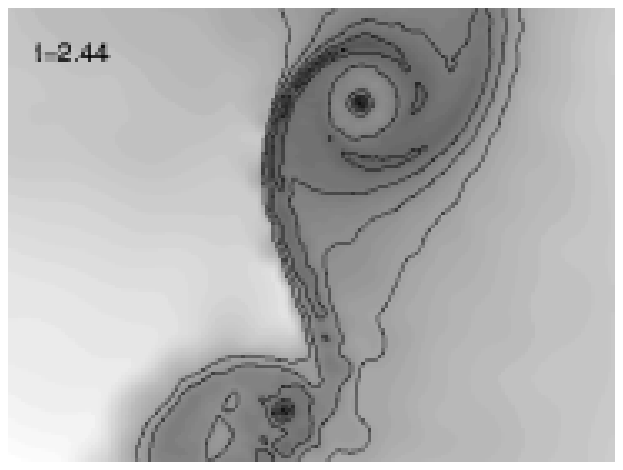}{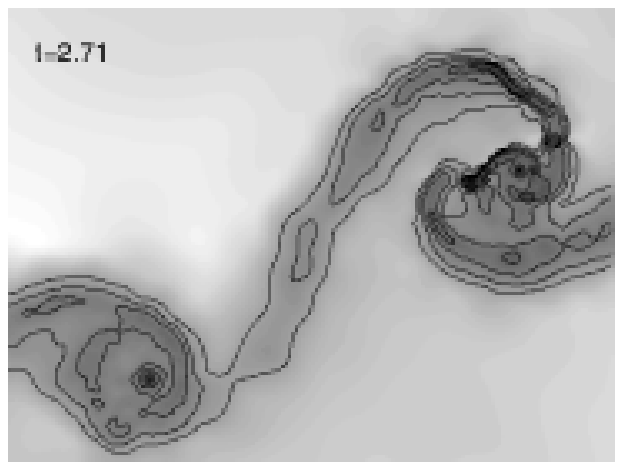}\\
\vspace{6pt}
\plottwo{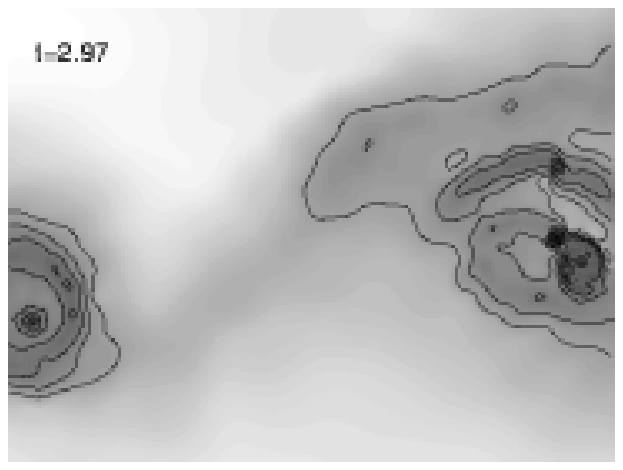}{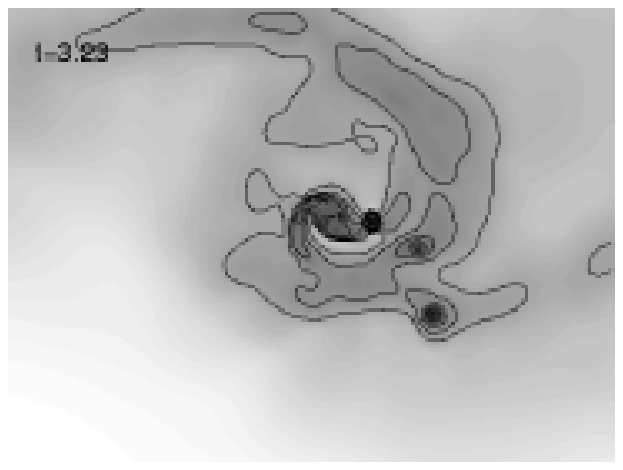}\\
\caption[Simulation ds02 at higher resolution (11301 ptcls): $\phi=0$, $r_{peri}=1000 \mbox{AU}$. 2000 $\times$ 1500 AU region.]{\label{fig:hires}Simulation ds02 at higher resolution (11301 ptcls): $\phi=0$, $r_{peri}=1000 \mbox{AU}$. 2000 $\times$ 1500 AU region. Contour levels at [0.3, 1, 3, 9, 30, 300] $\mbox{g cm}^{-2}$. The perturber approaches from the right-hand side. The 
evolution is clearly similar to the lower resolution one (Fig. 1). The end state consists, however, of 4 (maybe 5) stars instead of 3.} 
\end{figure*}

We have rerun simulation ds02 with more particles to 
see the effect of increasing the resolution. The disc is now made of 11299 particles.
The results are shown in Fig.~\ref{fig:hires}. Comparing this
with Fig. 1, we see that the main features in the two runs are similar,
in particular the appearance of the spiral arms and the fragmentation of the disc
around the primary. By the end of the simulation, the secondary had accreted 
$0.09 \mbox{M}_{\sun}$. Because the resolution is better, however, we are able
to resolve 2 or maybe 3 companions around the primary. Their masses are $0.117 \mbox{M}_{\sun}$,
$0.023 \mbox{M}_{\sun}$ and $0.007 \mbox{M}_{\sun}$. The primary is now surrounded by a disc
of only $0.0325 \mbox{M}_{\sun}$.
Thus, due to the higher resolution, mass which was still in a disc around the primary
in the low resolution run, has been accreted onto the newly formed close companion. In addition, two other less massive companions have formed from the disc. 
From this, we conclude that our lower resolution runs still capture the
important features of the flow, and that with increased resolution, the genesis of new protostellar
discs would be even more efficient.

\section{Vertically resolved disc}

As another check of the physical validity of the fragmentation we obtain, we 
have rerun simulation ds02 with a vertically resolved disc.
To make such a disc, we duplicated the evolved, single-layer disc, seven times and superposed the duplicates in the vertical direction so 
as to ensure that the disc was in vertical hydrostatic equilibrium. 
This vertically resolved disc was then evolved in isolation to remove
any transients. The way the disc was constructed, however, insured that it
quickly settled into a stable state.
The vertically resolved disc was composed of 13713 particles.

\begin{figure*}
\plottwo{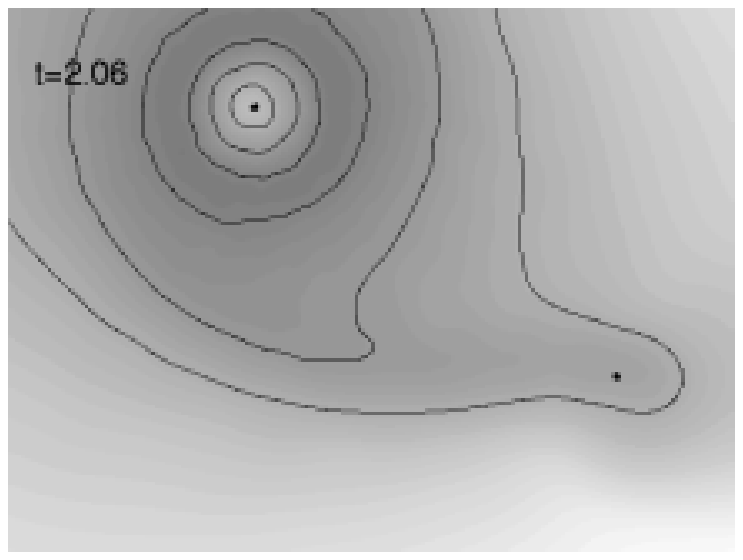}{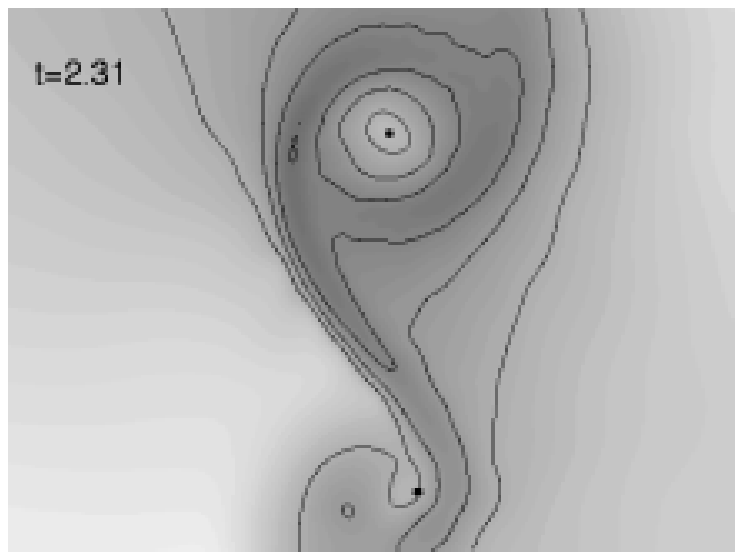}\\
\vspace{6pt}
\plottwo{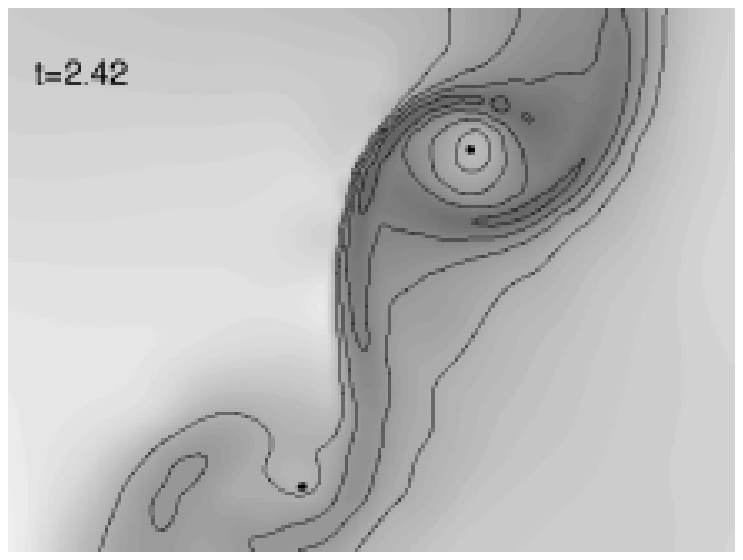}{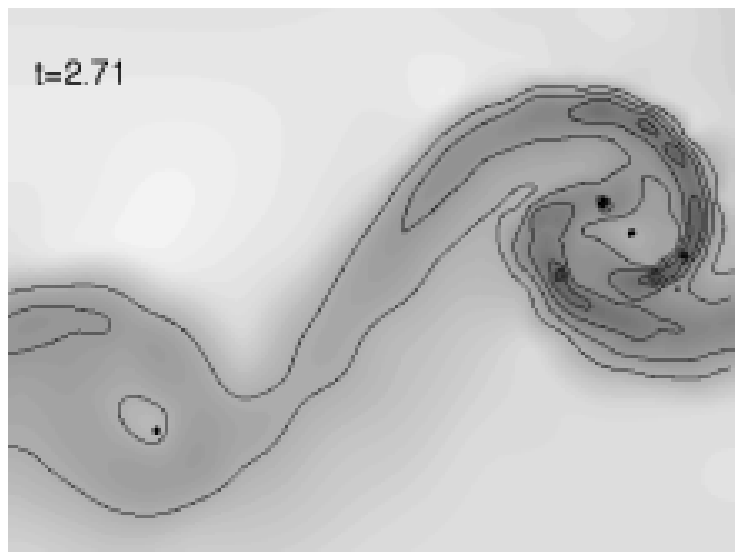}\\
\vspace{6pt}
\plottwo{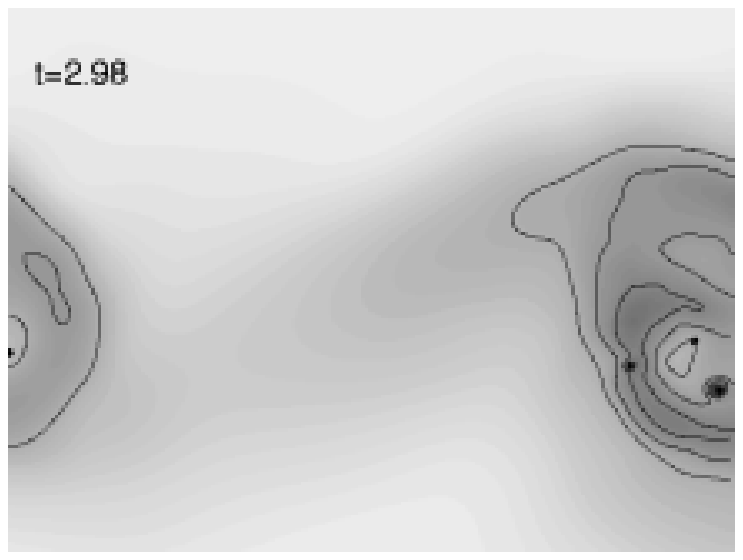}{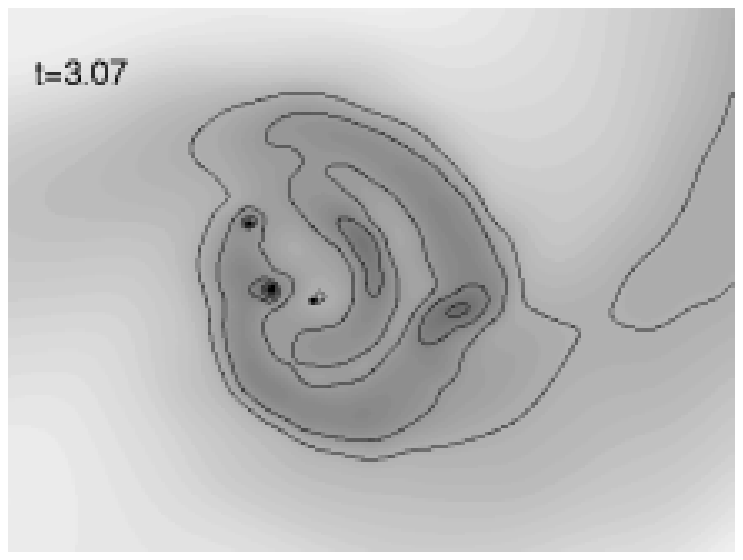}\\
\caption[Simulation ds02 using a vertically resolved disc (13715 ptcls): $\phi=0$, $r_{peri}=1000 \mbox{AU}$. 2000 $\times$ 1500 AU region.]{\label{fig:ds023d}Simulation ds02 using a vertically resolved disc (13715 ptcls): $\phi=0$, $r_{peri}=1000 \mbox{AU}$. 2000 $\times$ 1500 AU region. Contour levels at [0.3, 1, 3, 9, 30, 300] $\mbox{g cm}^{-2}$. The perturber approaches from the right-hand side. The 
evolution is similar to the lower resolution one (Fig. 1). The end state consists of 4 stars. (Note that the end state of this simulation is a little earlier than for the corresponding two-dimensional simulations illustrated in Figures 1 and B1.)} 
\end{figure*}

The results of this simulation are shown in figure \ref{fig:ds023d}.
Again, we see 
that the main features of Fig. 1 (the morphology of the spiral arms, and fragmentation of the
disc around the primary) are essentially the same.
After 2.58 10$^4$ years, two protostars start to form around the primary. 
Two other protostars form 1300 years later, one of which is directly destroyed by the primary
while the other merges with one formed previously.

By the end of the simulation (i.e. after 3.07 10$^4$ years), the secondary had accreted 
$0.07 \mbox{M}_{\sun}$. 
The two companions around the primary had masses of $0.033 \mbox{M}_{\sun}$ and
$0.016 \mbox{M}_{\sun}$. The primary was surrounded by a disc
of $0.13 \mbox{M}_{\sun}$. These values are comparable with those obtained in the standard ds02 simulation reported in Section 4.1. The differences are attributable to different seed noise in the different simulations.

\begin{thebibliography}{}
  \bibitem[\protect\citename{Bate et al.\ }1995]{bate} Bate M. R., Bonnell I. A., Price N. M., 1995, MNRAS, 277, 362
  \bibitem[\protect\citename{Beckwith \& Sargent }1993]{beckwith:sargent} Beckwith S. V. W., Sargent A. I., 1993, ApJ, 402, 280
  \bibitem[\protect\citename{Binney \& Tremaine }1987]{binney:tremaine} Binney J., Tremaine S., 1987, Galactic Dynamics, Princeton Univ. Press, Princeton
  \bibitem[\protect\citename{Chapman et al.\ }1992]{chapman92} Chapman S., Pongracic H., Disney M., Nelson A., Turner J., Whitworth A., 1992, Nat, 359, 207
  \bibitem[\protect\citename{Clarke \& Pringle }1991a]{clarke:pringle91a} Clarke C. J., Pringle J. E., 1991a, MNRAS, 249, 584
  \bibitem[\protect\citename{Clarke \& Pringle }1991b]{clarke:pringle91b} Clarke C. J., Pringle J. E., 1991b, MNRAS, 249, 588
  \bibitem[\protect\citename{Clarke \& Pringle }1993]{clarke:pringle93} Clarke C. J., Pringle J. E., 1993, MNRAS, 261, 190 (CP93)
  \bibitem[\protect\citename{Elmegreen \& Lada }1977]{elmegreen:lada} Elmegreen B. G., Lada C. J., 1977, ApJ, 214, 725
  \bibitem[\protect\citename{Hall et al.\ }1995]{hall} Hall S. M., Clarke C. J., Pringle J. E., 1995, MNRAS, 278, 303 (HCP96)
  \bibitem[\protect\citename{Heller }1993]{heller} Heller C. H., 1993, ApJ, 408, 337
  \bibitem[\protect\citename{Heller }1995]{heller95} Heller C. H., 1995, ApJ, 455, 252
  \bibitem[\protect\citename{Hernquist \& Katz }1989]{hernquist:katz} Hernquist L., Katz N., 1989, ApJSS, 70, 419
  \bibitem[\protect\citename{Korycansky \& Papaloizou }1995]{korycansky:papaloizou} Korycansky D. G., Papaloizou J. C., 1995, MNRAS, 274, 85
  \bibitem[\protect\citename{Lada et al.\ }1991]{lada91} Lada E. A., DePoy D. L., Evans N. J. II, Gatley I., 1991, ApJ, 371, 171
  \bibitem[\protect\citename{Larson }1990]{larson90} Larson R. B., 1990, in Capuzzo-Dolcetta R., Chiosi C., Di Fazio A., eds, Physical processes in fragmentation and star formation
  \bibitem[\protect\citename{Larwood et al.\ }1996]{larwood} Larwood J. D., Nelson R. P., Papaloizou J. C. B., Terquem C., 1996, MNRAS, 282, 597
  \bibitem[\protect\citename{Laughlin \& Rozyczka }1996]{laughlin:rozyczka} Laughlin G., Rozyczka M., 1996, ApJ, 456, 279
  \bibitem[\protect\citename{Lin \& Pringle }1990]{lin:pringle90} Lin D. C., Pringle J. E., 1990, ApJ, 358, 515
  \bibitem[\protect\citename{Lynden-Bell \& Pringle }1974]{lbell:pring} Lynden-Bell D., Pringle, J. E., 1974, MNRAS, 168, 603
  \bibitem[\protect\citename{McDonald \& Clarke }1995]{mcdonald:clarke} McDonald J. M., Clarke C. J., 1995, MNRAS, 275, 671
  \bibitem[\protect\citename{Mathieu }1994]{mathieu94} Mathieu R. D., 1994, ARA\&A, 32, 465
  \bibitem[\protect\citename{Monaghan }1989]{monaghan89} Monaghan J. J., 1989, J. Comput. Phys., 82, 1
  \bibitem[\protect\citename{Monaghan }1992]{monaghan92} Monaghan J. J., 1992, ARA\&A, 30, 543
  \bibitem[\protect\citename{Murray }1996]{murray} Murray J. R., 1996, MNRAS, 279, 402
  \bibitem[\protect\citename{Ostriker }1994]{ostriker} Ostriker E. C., 1994, ApJ, 424, 292
  \bibitem[\protect\citename{Press et al.\ }1992]{press92} Press W. H., Teukolsky S. A., Vetterling W. T., Flannery B. P., 1992, Numerical Recipes in Fortran, 2nd edn., Cambridge Univ. Press, Cambridge
  \bibitem[\protect\citename{Pringle }1989]{pringle89} Pringle J. E., 1989, MNRAS, 239, 361
  \bibitem[\protect\citename{Rasio \& Shapiro }1991]{ras:sha} Rasio, F.A., Shapiro, S.L., 1991, ApJ, 377, 559
  \bibitem[\protect\citename{Reipurth \& Zinnecker }1993]{reipurth} Reipurth B., Zinnecker H., 1993, A\&A, 278, 81
  \bibitem[\protect\citename{Shakura \& Sunyaev }1973]{shakura:sunyaev} Shakura N. I., Sunyaev R. A., 1973, A\&A, 24, 337
  \bibitem[\protect\citename{Sod }1978]{sod} Sod G. A., 1978, J. Comput. Phys., 27, 1
  \bibitem[\protect\citename{Strom, Margulis \& Strom }1989a]{strom89a} Strom K. M., Margulis M., Strom S. E., 1989a, ApJ, 346, L33
  \bibitem[\protect\citename{Strom et al.\ }1989b]{strom89b} Strom K. M., Strom S. E., Edwards S., Cabrit S., Skrutskie M. F., 1989b, AJ, 97, 1451
  \bibitem[\protect\citename{Terquem \& Bertout }1993]{terquem:bertout93} Terquem C., Bertout C., 1993, A\&A, 274, 291
  \bibitem[\protect\citename{Terquem \& Bertout }1996]{terquem:bertout96} Terquem C., Bertout C., 1996, MNRAS, 279, 415
  \bibitem[\protect\citename{Toomre  }1964]{toomre} Toomre A., 1964, ApJ, 139, 1217
  \bibitem[\protect\citename{Toomre \& Toomre }1972]{toomre:toomre} Toomre A., Toomre J., 1972, ApJ, 178, 623
\bibitem[\protect\citename{Turner et al.\  }1995]{turner} Turner J. A., Bhattal A. S., Chapman S. J., Disney M. J., Pongracic H., Whitworth A. P., 1995, MNRAS, 277, 705
  \bibitem[\protect\citename{Watkins et al.\ }1996]{watkins} Watkins S. J., Bhattal A. S., Francis N., Whitworth, A. P., 1996, A\&ASS, 119, 177
  \bibitem[\protect\citename{Watkins et al.\ }1997b]{watkins97b} Watkins S. J., Bhattal A. S., Boffin H. M. J., Francis N., Whitworth, A. P., 1997b, MNRAS, submitted
  \bibitem[\protect\citename{Watkins et al.\ }1997c]{watkins97c} Watkins S. J., Bhattal A. S., Boffin H. M. J., Francis N., Whitworth, A. P., 1997c, MNRAS, submitted
\bibitem[\protect\citename{Whitworth et al.\  }1995]{whitworth} Whitworth A. P., Bhattal A. S., Chapman S. J., Disney M. J., Pongracic H., Turner J. A., 1995, MNRAS, 277, 727
\bibitem[\protect\citename{Whitworth et al.\  }1996]{antisis} Whitworth A. P., Bhattal A. S., Francis N., Watkins S. J., 1996, MNRAS, 283, 1061
\end{thebibliography}
\end{document}